\documentclass[12pt]{article}
\usepackage{comment}
\usepackage{amsmath}
\usepackage{amssymb}
\usepackage{graphicx}
\usepackage{here}
\usepackage{subcaption}
\usepackage{url}
\usepackage[sort&compress, numbers, merge]{natbib}
\usepackage{physics}
\usepackage{todonotes}

\newcommand{\Slash}[1]{{\ooalign{\hfil/\hfil\crcr$#1$}}}

\usepackage[colorlinks=true, linkcolor=blue, citecolor=blue, urlcolor=black]{hyperref}

\begin{document}

\def\ps{\mathbf{p}}
\def\PS{\mathbf{P}}

\baselineskip 0.6cm

\def\simgt{\mathrel{\lower2.5pt\vbox{\lineskip=0pt\baselineskip=0pt
           \hbox{$>$}\hbox{$\sim$}}}}
\def\simlt{\mathrel{\lower2.5pt\vbox{\lineskip=0pt\baselineskip=0pt
           \hbox{$<$}\hbox{$\sim$}}}}
\def\simprop{\mathrel{\lower3.0pt\vbox{\lineskip=1.0pt\baselineskip=0pt
             \hbox{$\propto$}\hbox{$\sim$}}}}
\def\tr{\mathop{\rm tr}}
\def\SU{\mathop{\rm SU}}

\begin{titlepage}

\begin{flushright}
IPMU20-0125
\end{flushright}
~

\vskip 1.5cm

\begin{center}

{\Large \bf
Non-perturbative Effects on Electroweakly Interacting \\
\vspace{0.2cm} Massive Particles at Hadron Collider}

\vskip 2cm
{\large
Taisuke Katayose, Shigeki Matsumoto and Satoshi Shirai
}

\vskip 1.0cm
{\it
Kavli Institute for the Physics and Mathematics of the Universe (WPI), \\
The University of Tokyo Institutes for Advanced Study, \\
The University of Tokyo, Kashiwa 277-8583, Japan
}

\vskip 3.5cm
\abstract{
Electroweakly Interacting Massive Particles (EWIMPs), in other words, new massive particles that are charged under the electroweak interaction of the Standard Model (SM), are often predicted in various new physics models. EWIMPs are probed at hadron collider experiments not only by observing their direct productions but also by measuring their quantum effects on Drell-Yan processes for SM lepton pair productions. Such effects are known to be enhanced especially when the di-lepton invariant mass of the final state is close to the EWIMP threshold, namely twice the EWIMP mass. In such a mass region, however, we have to carefully take non-perturbative effects into account, because the EWIMPs become non-relativistic and the prediction may be significantly affected by e.g., bound states of the EWIMPs caused by the electroweak interaction. We study such non-perturbative effects using the non-relativistic effective field theory of the EWIMPs, and found that those indeed affect the differential cross section of the Drell-Yan processes significantly, though the effects are smeared due to the finite energy resolution of the lepton measurement at the Large Hadron Collider experiment.
}

\end{center}

\end{titlepage}

\section{Introduction}
\label{sec:Introduction}

Many new physics models predict electroweakly interacting massive particles (EWIMPs), namely new massive particles that are charged under the $\mathrm{SU}(2)_L \times \mathrm{U}(1)_Y $ gauge interaction of the standard model (SM). One example is the extension of the Higgs sector, where new scalar particles carrying various $\mathrm{SU}(2)_L \times \mathrm{U}(1)_Y$ charges are usually introduced. Another example is the supersymmetry or extra-dimension scenario, where many copies of the SM particles are predicted in general.

The EWIMP is also known to become a good candidate for dark matter\,\cite{Cirelli:2005uq, *Cirelli:2007xd, *Cirelli:2009uv, Nagata:2014aoa, Abe:2020mph}. Then, the electroweak interaction plays an essential role in the freeze-out mechanism, which leads to the dark matter abundance, $\Omega h^2 \sim 0.1$, naturally. Concrete examples of the EWIMP dark matter are as follows: In the minimal dark matter model\,\cite{Cirelli:2005uq, *Cirelli:2007xd, *Cirelli:2009uv}, a large electroweak charge automatically stabilizes the EWIMP to live long enough as dark matter without imposing any ad-hoc symmetry. On the other hand, many new physics models concerning the electroweak symmetry breaking also predict the EWIMP dark matter. For instance, many supersymmetric standard models predict the Higgsino (doublet quasi-Dirac fermion) or wino (triplet Majorana fermion) as the lightest supersymmetric particle, namely dark matter. In particular, the wino dark matter is known to be the most well-motivated: The dark matter is the prediction of the anomaly mediation\,\cite{Randall:1998uk, Giudice:1998xp}, where this framework attracts great attention after the discovery of the Higgs boson and it stimulates various model building\,\cite{Hall:2011jd, Hall:2012zp, Nomura:2014asa, Ibe:2011aa, Ibe:2012hu, Arvanitaki:2012ps, ArkaniHamed:2012gw} as well as phenomenological studies of the wino dark matter\,\cite{Hisano:2003ec, Hisano:2004ds, Hisano:2005ec, Hisano:2010fy, Hisano:2010ct, Hisano:2012wm, Hisano:2015rsa}.

The search of the EWIMP is therefore the most important task of collider experiments. The conventional search strategy strongly relies on how the EWIMP decays, especially, the $Q$-value. Generally, if the mass difference between the EWIMP and daughter particle(s) is smaller, the decay products are less energetic and the sensitivity of the collider search gets weaker. In the case of the EWIMP dark matter, mass differences among the $\mathrm{SU}(2)_{L}$ multiplet are small\,\cite{Yamada:2009ve, Ibe:2012sx, McKay:2017rjs, Nagata:2014wma, Nagata:2014aoa}, and accordingly the EWIMP dark matter becomes a tough target at hadron collider experiments, regardless of their large production cross sections. In order to probe the EWIMP dark matter, we need special signatures such as a disappearing charged track\,\cite{Ibe:2006de, Buckley:2009kv, Asai:2007sw, Asai:2008sk, Asai:2008im, Mahbubani:2017gjh, Fukuda:2017jmk, Saito:2019rtg, Chigusa:2019zae} or soft track\,\cite{Fukuda:2019kbp} which originates in a meta-stable charged EWIMP component. Improved detection of such exotic signatures is thus essential for the discovery of the EWIMP and is now intensively being studied in many literature. 

We have proposed a new strategy of the EWIMP search at high-energy colliders which does not rely on the decay of the EWIMP. This method utilizes the indirect effect of EWIMP to SM processes\,\cite{Harigaya:2015yaa, Matsumoto:2017vfu, Matsumoto:2018ioi}. As the EWIMP has an electroweak charge by its definition, it gives quantum corrections to the self-energies of electroweak gauge bosons. We are therefore able to observe the indirect signature from the precision measurement of appropriate SM processes, such as di-fermion productions. As this method does not assume the decay of the EWIMP, it is also possible to provide the most conservative constraint on the EWIMP. For hadron colliders, Drell-Yan processes, $pp\to W(\gamma/Z) X \to \ell \bar{\nu} (\ell \bar{\ell}) X$, give the most sensitive probe. There are several studies along this line for the Large Hadron Collider (LHC) and future hadron collider experiments\,\cite{Chigusa:2018vxz, DiLuzio:2018jwd, Abe:2019egv}.

Previous studies have discussed the one-loop effect of this correction. The strongest signature of the EWIMP in the Drell-Yan processes appears, when the invariant mass of the final state lepton pair is around the twice of the EWIMP mass, $m_{\ell \ell} \simeq 2 M_\mathrm{EWIMP}$. In this threshold region, however, the virtual EWIMP is almost on-shell and non-relativistic, and the one-loop analysis is no longer validated. For instance, two-loop diagram has a mass-threshold singularity, and non-perturbative bound-states and Sommerfeld effects become rather significant. We compute self-consistent higher order corrections of the EWIMP in this article, and discuss its impact on the collider studies.

\section{EWIMP oblique correction to Drell-Yan processes}
\label{sec: oblique correction}

In this article, we focus on quantum corrections from electroweakly interacting massive particles (EWIMPs) on Drell-Yan processes for standard model (SM) lepton pair productions at hadron collider experiments such as Large Hadron Collider (LHC). The most important correction comes from the so-called oblique correction to self-energies of electroweak gauge bosons shown below:

\begin{figure}[h!] 
    \centering
    \includegraphics[width=0.7\linewidth]{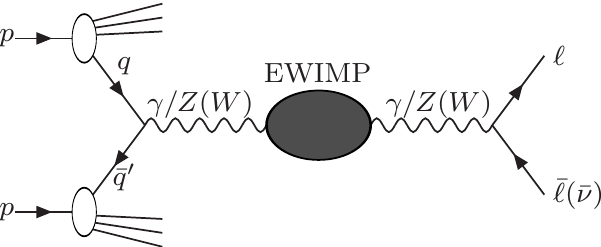} 
\end{figure}

\noindent
The oblique correction is expressed by a filled circle above, and parameterized by $\Delta \Pi^{\mu\nu}$ as follows:
\begin{align}
    \Delta \Pi^{\mu\nu}_{V V'} =  q^2\,\Pi_{V V'}(q^2)\,g^{\mu\nu} + \Sigma_{V V'}(q^2)\,{q^\mu q^\nu} ,
\end{align}
where $V$, $V'$ denote $\gamma$, $Z$, $W$ bosons, and $q_\mu$ is the momentum of the propagating electroweak gauge bosons. Since external fermion masses are negligibly small, we ignore the $\Sigma_{V V'}$ part hereafter.

Matrix elements of the Drell-Yan processes for SM lepton pair productions (the neutral and  charged current cases, respectively) at leading order calculation are given by following formulae:
\begin{align}
    \mathcal{M}^\mathrm{Neutral}_\mathrm{LO} [q(p)\bar{q}(p') \to \ell^-(k) \ell^+(k')] &=
    \sum_{V=\gamma,Z} \frac{[\bar{v}(q;p') \gamma^\mu \Gamma^V_q u(q;p)]\,[\bar{u}(\ell;k) \gamma_\mu \Gamma^V_\ell v(\ell;k')]}{\hat{s} - m_V^2} , 
    \label{eq:neutral_LO} \\
	\mathcal{M}^\mathrm{Charged}_\mathrm{LO} [u(p)\bar{d}(p') \to \ell^+(k) {\nu}(k')] &=
	\frac{g^2}{2} \frac{[\bar{v}_L(d;p') \gamma^\mu  u_L(u;p)]\,[\bar{u}_L(\ell;k) \gamma_\mu v_L(\nu;k')]}{\hat{s} - m^2_W} ,
	\label{eq:charged_LO}
\end{align}
where $\hat{s}$ is the center-of-mass energy squared at each parton-level process, while $\Gamma^Z_f = g_Z (v_f - a_f \gamma_5)$ and $\Gamma^\gamma_f = e\,Q_f$, where $g_Z \equiv (g^2+g'^2)^{1/2}$ and $e \equiv g g'/(g^2+g'^2)^{1/2}$ with $g$ and $g'$ being $\mathrm{SU}(2)_L$ and $\mathrm{U}(1)_Y$ gauge couplings, respectively. The electric charge of the fermion `$f$' is denoted by $Q_f$, while $v_f = T^3_f/2 - Q_f \sin^2\theta_W$ and $a_f = T^3_f/2$ with $T^3_f$ being the weak-charge of `$f$', and $\sin\theta_W \equiv g'/g_Z$. The matrix elements discussed above are corrected by EWIMPs, and those are given by
\begin{align}
	\mathcal{M}^\mathrm{Neutral}_\mathrm{EWIMP} [q(p)\bar{q}(p') \to \ell^-(k) \ell^+(k')] &= 
	\sum_{V, V'} \frac{[\bar{v}(q;p') \gamma^\mu \Gamma^V_q u(q;p)]
	\,\hat{s}\,\Pi_{V V'}(\hat{s})\,
	[\bar{u}(\ell;k) \gamma_\mu \Gamma^{V'}_\ell v(\ell;k')]}
	{ (\hat{s} - m^2_V) (\hat{s} - m^2_{V'}) } ,
	\label{eq:neutral_correction} \\
	\mathcal{M}^\mathrm{Charged}_\mathrm{EWIMP} [u(p)\bar{d}(p') \to \ell^+(k) {\nu}(k')] &=
	\frac{g^2}{2} \, \frac{[\bar{v}_L(d;p') \gamma^\mu u_L(u;p)]
	\,\hat{s}\,\Pi_{W W}(\hat{s})\,
    [\bar{u}_L(\ell;k) \gamma_\mu v_L(\nu;k')]}{(\hat{s} - m^2_W)^2} .
	\label{eq:charged_correction}
\end{align}

Since the interference between the SM matrix element (LO matrix element), $\mathcal{M}_\mathrm{LO}$, and those of EWIMP loop contributions, $\mathcal{M}_\mathrm{EWIMP}$, give the largest correction to the Drell-Yan processes, the correction is roughly given by the aforementioned EWIMP oblique correction as follows:
\begin{align}
    \frac{ |\mathcal{M}_{\rm SM} + \mathcal{M}_{\rm EWIMP}|^2 - |\mathcal{M}_{\rm SM}|^2 }{ |{\cal M}_{\rm SM}|^2 }
    \sim 2 {\rm Re}[\Pi_{V V}(\hat{s})] ,
    \label{eq:correction}
\end{align}
with $V$ being $W^\pm$ for the charged current case, and $W^3$, namely the neutral component of the weak gauge boson multiplet, for the neutral current case when the EWIMP does not carry a hypercharge. In the following two sections, we discuss the EWIMP correction to the self-energies of the electroweak gauge bosons at one-loop, two-loop and non-perturbative orders. We focus on the case of the wino-like particle ($\mathrm{SU}(2)_L$ triplet Majorana fermion) in this article to make our discussion concrete. 

\section{Perturbative corrections}
\label{sec: perturbative corrections}

We consider the $\mathrm{SU}(2)_L$ triplet Majorana fermion $\chi$ throughout this paper as the simplest example of the EWIMP. The corresponding new physics Lagrangian is then simply given as follows:
\begin{eqnarray}
    \mathcal{L} = \mathcal{L}_{\mathrm{SM}}
    + i \chi^\dagger D_\mu \sigma^{\mu} \chi
    - \frac{M_\chi}{2}(\chi^T \epsilon \chi - \chi^\dagger \epsilon \chi^*) ,
    \label{eq: Lagrangian}
\end{eqnarray}
where $\mathcal{L}_{\mathrm{SM}}$ is the SM Lagrangian, $M_\chi$ is the EWIMP mass, and $D_\mu \equiv \partial_\mu - i g T^a W^a_\mu$ is the covariant derivative acting on the EWIMP with $g$, $T^a$, $W^a_\mu$ being gauge coupling, generator, gauge boson field of the $\mathrm{SU}(2)_L$ interaction, respectively.
The EWIMP field $\chi$ is composed of three Weyl fermions,
\begin{eqnarray}
    \chi &\equiv& (\chi^+, \chi^0, \chi^- )^T .
\end{eqnarray}

With neutral and charged components of the EWIMP field, the leading perturbative correction to the self-energies of the electroweak gauge bosons is obtained by calculating one-loop diagrams,
\begin{figure}[H] 
    \centering
    \includegraphics[width=58mm]{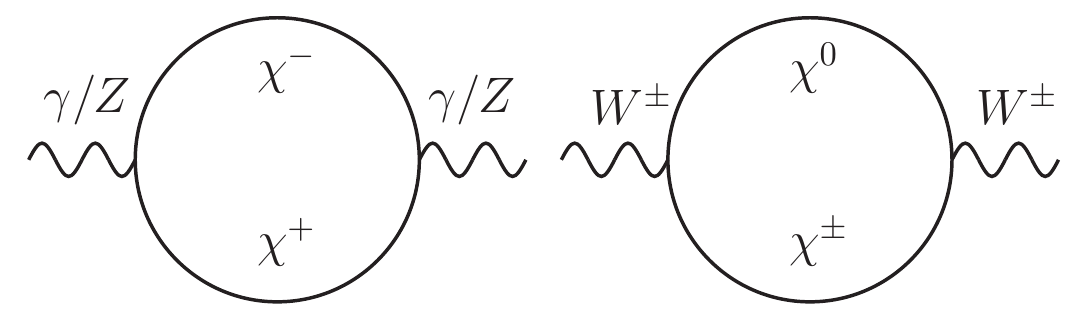}
\end{figure}
\noindent
and EWIMP oblique correction to the electroweak gauge bosons up to one-loop level is given by
\begin{align}
    & \Pi_{W W}(q^2) = \Pi_{\mathrm{1\mathchar`-loop}}(q^2/m_{\chi}^2) , \\
    & \Pi_{\gamma \gamma}(q^2) = \sin^2\theta_W\, \Pi_{\mathrm{1\mathchar`-loop}}(q^2/m_\chi^2) , \\
    & \Pi_{\gamma Z}(q^2) = \sin\theta_W \cos\theta_W\,\Pi_{\mathrm{1\mathchar`-loop}}(q^2/m_\chi^2) , \\
    & \Pi_{Z Z}(q^2) = \cos^2\theta_W\,\Pi_{\mathrm{1\mathchar`-loop}}(q^2/m_\chi^2) ,
    \label{eq: One-loop oblique correction}
\end{align}
where $m_\chi =m_{\chi^\pm} = m_{\chi^0}$ being physical masses of the charged and neutral components of the EWIMP.
Here, we neglect the small mass difference between $\chi^\pm$ and $\chi^0$.
The function $\Pi_{\mathrm{1\mathchar`-loop}}(x^2)$ appearing in the above formula is expressed by a simple integration-form, as addressed in many past literature such as Ref.\,\,\cite{Matsumoto:2017vfu}:
\begin{eqnarray}
   \Pi_{\mathrm{1\mathchar`-loop}}(x^2) \equiv \frac{g^2}{2\pi^2} \int_0^1 dy\,y(1-y)\mathrm{ln}[1 - y(1-y)x^2 - i0^+] .
\end{eqnarray}
Here, we take the dimensional regularization to regularize one-loop integrals and fix the renormalization condition for gauge boson kinetic terms so that the correction vanishes at $q^2 = 0$.

\begin{figure}[t] 
    \centering
    \includegraphics[width=82mm]{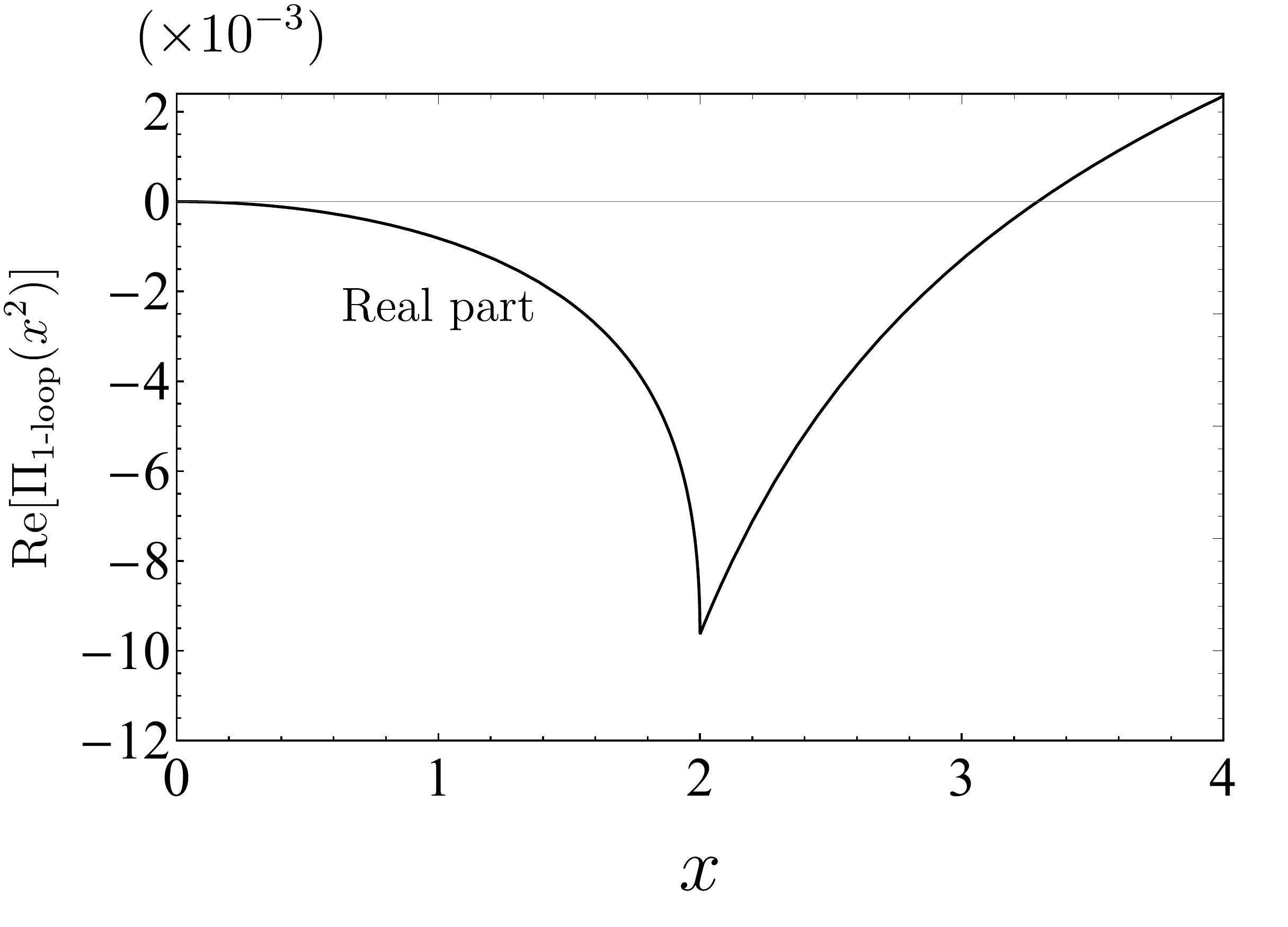}
    \qquad
    \includegraphics[width=82mm]{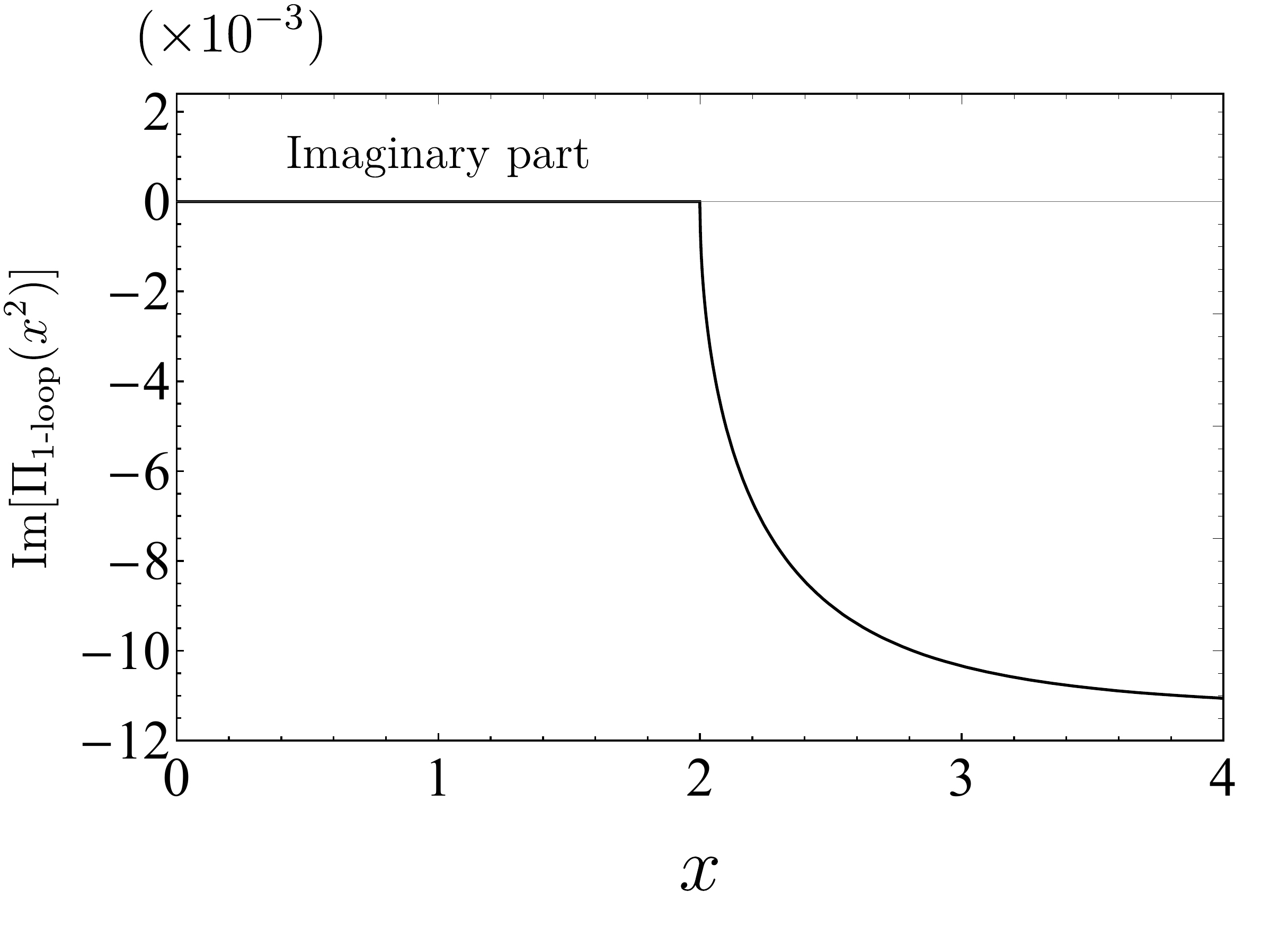}
    \caption{\small Real and imaginary parts of the leading perturbative correction to the self-energy of the electroweak gauge bosons, $\Pi_{\mathrm{1\mathchar`-loop}}(x^2)$, obtained by calculating the one-loop diagrams in the text.}
    \label{fig: 1loop}
\end{figure}

The real part of the leading perturbative correction, $\Re[\Pi_{\mathrm{1\mathchar`-loop}}(x^2)]$, has a cusp structure at the threshold ($x = 2$) as seen in Fig.\,\ref{fig: 1loop}, and it takes the value of $- 2g^2/(9\pi^2)$ there. Hence, di-lepton production cross sections of the Drell-Yan processes are reduced at the threshold region, namely the cross sections have a negative bump structure at $m_{\ell\ell} \simeq 2m_{\chi}$ with $m_{\ell\ell}$ being the di-lepton invariant mass. It thus allows us to search for the EWIMP by measuring $m_{\ell\ell}$ distribution precisely. It should however be noted that the EWIMP becomes non-relativistic at the threshold region, and usual perturbative expansion is not guaranteed to work to compute the oblique correction due to the threshold singularity. In the next section, we will discuss non-perturbative contributions to the correction using the non-relativistic (NR) Lagrangian method. Before going to this discussion, on the other hand, we also discuss the next-leading perturbative correction to the correction in the rest of this section, for it enables us to smoothly match the result from the NR Lagrangian method with the one from the perturbative method between in- and outside the threshold region.

The next-leading perturbative correction to the self-energies of the electroweak gauge bosons is obtained by calculating two-loop diagrams in which EWIMP and electroweak gauge bosons are propagating in the loops. Among various diagrams, those relevant to our discussion are
\begin{figure}[H]
    \centering
    \includegraphics[width=82mm]{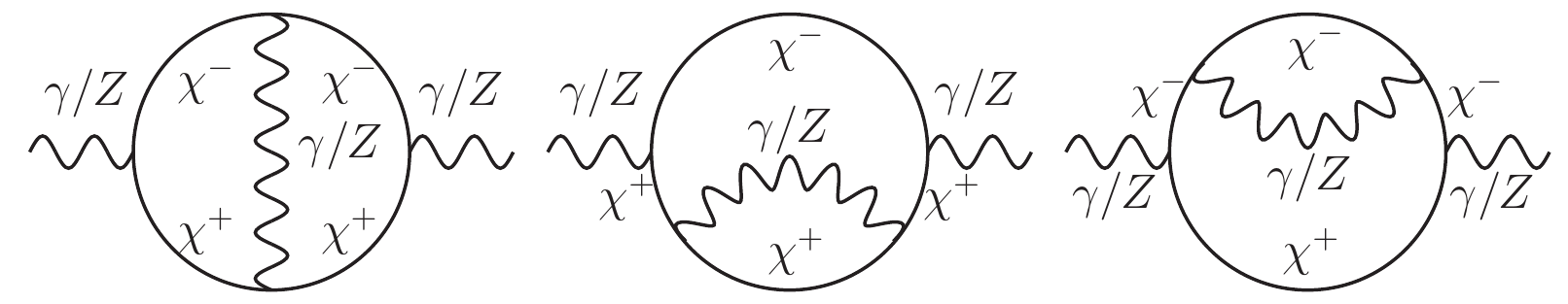}
    \qquad
    \includegraphics[width=82mm]{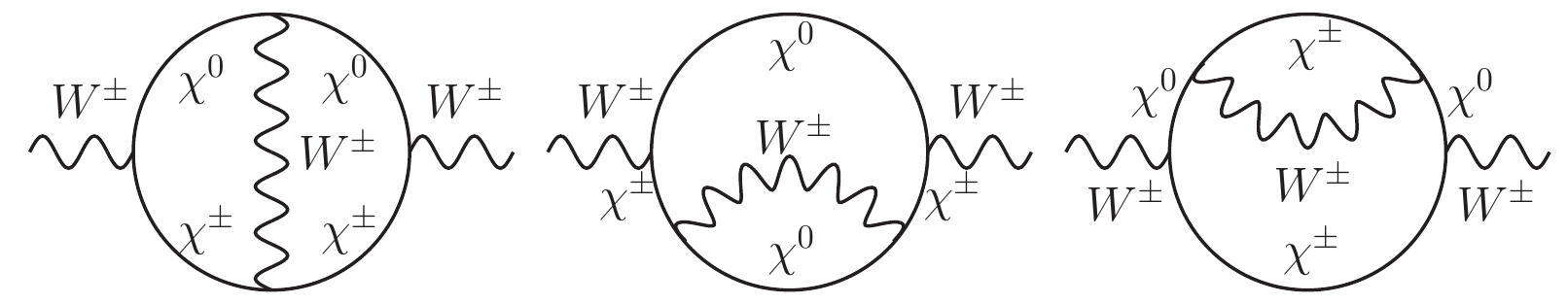}
\end{figure}
\noindent
where first and fourth diagrams become dominant at the threshold region, and those are also taken into account in the NR Lagrangian discussed in the next section. On the other hand, we also have to take the other diagrams (the second, third, fifth and sixth ones) into account to evaluate the next-leading perturbative correction, as these contributions remove the infrared singularity caused by the dominant diagrams. In addition to the above diagrams, there are other two-loop diagrams which originate in the non-Abelian nature of the $\mathrm{SU}(2)_L$ interaction. Since those are expected to contribute to the matching between the perturbative correction and the non-perturbative correction (obtained by the NR Lagrangian) sub-dominantly, we do not include those in our analysis.

Among various contributions to the next-leading perturbative correction in the above diagrams, those including a massless photon-exchange have already been computed analytically\,\cite{Hoang:1997vs},
\begin{align}
    \Pi_{\mathrm{2\mathchar`-loop};\,\gamma}^0 (x^2) \equiv -\frac{g^2\alpha}{4\pi^2}
    \Bigg\{
        & \frac{18-13\beta^2}{24} + \frac{\beta(5-3\beta^2)}{8} \ln{(-p)} - \frac{(1-\beta)(33-39\beta-17\beta^2+7\beta^3)}{96}[\ln{(-p)}]^2
        \nonumber \\ 
        & \frac{\beta(-3+\beta^2)}{3} \Big[ 2\ln{(1-p)}\ln{(-p)} + \ln{(-p)}\ln{(1+p)} + \mathrm{Li}_2(-p)+2\mathrm{Li}_2(p) \Big]
        \nonumber \\ 
        & +\frac{(3-\beta^2)(1+\beta^2)}{12} \Big[ 2\ln{(1-p)}[\ln{(-p)}]^2 + [\ln{(-p)}]^2\ln{(1+p)}
        \nonumber \\  
        & +4\ln{(-p)}\mathrm{Li}_2(-p) + 8\ln{(-p)}\mathrm{Li}_2(p) - 6\mathrm{Li}_3(-p) - 12\mathrm{Li}_3(p) -3\zeta_3 \Big]
    \Bigg\} ,
\end{align}
where $\alpha$ is the fine structure constant, $\mathrm{Li}_2$ and $\mathrm{Li}_3$ are the di- and tri-logarithms, respectively, while $\beta \equiv [1-4/(x^2 + i0^+)]^{1/2}$, $p \equiv (1-\beta)/(1+\beta)$ and $\zeta_3 \simeq 1.2020569$ with $\zeta_x$ being the Zeta function. The superscript `0' appearing at $\Pi_{\mathrm{2\mathchar`-loop};\,\gamma}^0$ means that this is the correction to the self-energy of the neutral weak gauge boson ($Z$ and $\gamma$). We will use the expression $\Pi_{\mathrm{2\mathchar`-loop}}^\pm$ to denote the next-leading perturbative correction to the self-energy of the charged weak gauge boson ($W^\pm$). Other contributions including a massive gauge boson-exchange, $\Pi_{\mathrm{2\mathchar`-loop};\,Z}^0$ and $\Pi_{\mathrm{2\mathchar`-loop};\,W}^\pm$, in the above two-loop diagrams are numerically evaluated using the codes TSIL\,\cite{Martin:2005qm} and TARCER\,\cite{Mertig:1998vk}.

To evaluate the next-leading perturbative corrections, we take the dimensional regularization to regularize the two-loop integrals and fix the same renormalization condition as those of the leading one for kinetic terms (two-point functions) of the electroweak gauge bosons. On the other hand, we also have to fix renormalization conditions at one-loop level for kinetic terms (two-point functions) of neutral and charged components of the EWIMP filed, and vertices (three-point functions) of neutral and charged current interactions of the field. The conditions we adopted are as follows:
\begin{align}
    & {\cal G}_{\chi^\pm}(p) = i/(\Slash{p} - m_{\chi}) \qquad {\rm when} \qquad p^2 = m_{\chi}^2 , \\
    & \Gamma_{\chi^+ \chi^- W^3}(p_1, p_2) = -i g \qquad {\rm when} \qquad p_1^2 = p_2^2 = m_{\chi}^2 ,
\end{align}
for the next-leading perturbative correction to the self-energy of the neutral electroweak gauge bosons. Here, ${\cal G}(p)$ is the propagator (two-point function) of the EWIMP field, while $\Gamma(p_1, p_2)$ is the vertex (three-point function) involving two EWIMP fields. On the other hand, the conditions for calculating the correction to the self-energy of the charged weak gauge boson are given by
\begin{align}
    & {\cal G}_{\chi^0}(p) = i/(\Slash{p} - m_\chi) \qquad {\rm when} \qquad p^2 = m_\chi^2 , \\
    & {\cal G}_{\chi^\pm}(p) = i/(\Slash{p} - m_\chi) \qquad {\rm when} \qquad p^2 = m_\chi^2 , \\
    & \Gamma_{\chi^0 \chi^\pm W^\mp}(p_1, p_2) = -i g \qquad {\rm when} \qquad p_1^2 = p_2^2 = m_\chi^2 .
\end{align}
It means that we take on-shell normalization conditions for the EWIMP, because it makes the convergence of the perturbative expansion proper even at the threshold region. With these corrections, EWIMP oblique correction to the electroweak gauge bosons up to two-loop level is give by
\begin{align}
    & \Pi_{W W}(q^2) = \Pi_{\mathrm{1\mathchar`-loop}}(q^2/m_{\chi}^2) + \Pi_{\mathrm{2\mathchar`-loop};\,W}^\pm(q^2/m_{\chi}^2) , \\
    & \Pi_{\gamma \gamma}(q^2) = \sin^2\theta_W\,[\, \Pi_{\mathrm{1\mathchar`-loop}}(q^2/m_\chi^2) + \Pi_{\mathrm{2\mathchar`-loop}}^0(q^2/m_\chi^2)\,] , \\
    & \Pi_{\gamma Z}(q^2) = \sin\theta_W \cos\theta_W\,[\, \Pi_{\mathrm{1\mathchar`-loop}}(q^2/m_\chi^2) + \Pi_{\mathrm{2\mathchar`-loop}}^0(q^2/m_\chi^2)\,] , \\
    & \Pi_{Z Z}(q^2) = \cos^2\theta_W\,[\, \Pi_{\mathrm{1\mathchar`-loop}}(q^2/m_\chi^2) + \Pi_{\mathrm{2\mathchar`-loop}}^0(q^2/m_\chi^2)\,] , \\
    & \Pi_{\mathrm{2\mathchar`-loop}}^0(x^2) \equiv \Pi_{\mathrm{2\mathchar`-loop};\,\gamma}^0(x^2) + \Pi_{\mathrm{2\mathchar`-loop};\,Z}^0(x^2) .
    \label{eq: Two-loop Oblique correction}
\end{align}

\begin{figure}[t] 
    \centering
    \includegraphics[width=82mm]{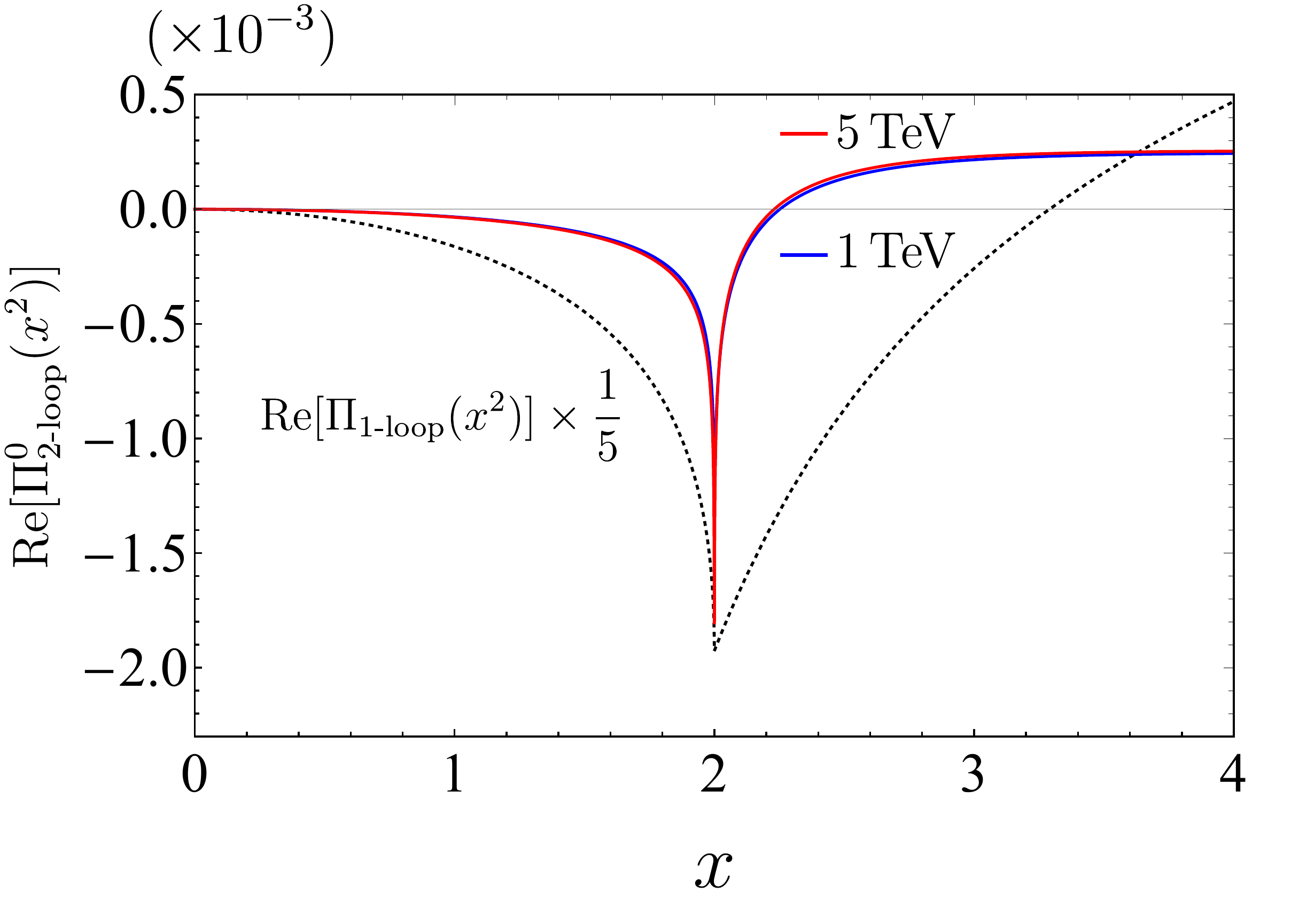}
    \qquad
    \includegraphics[width=82mm]{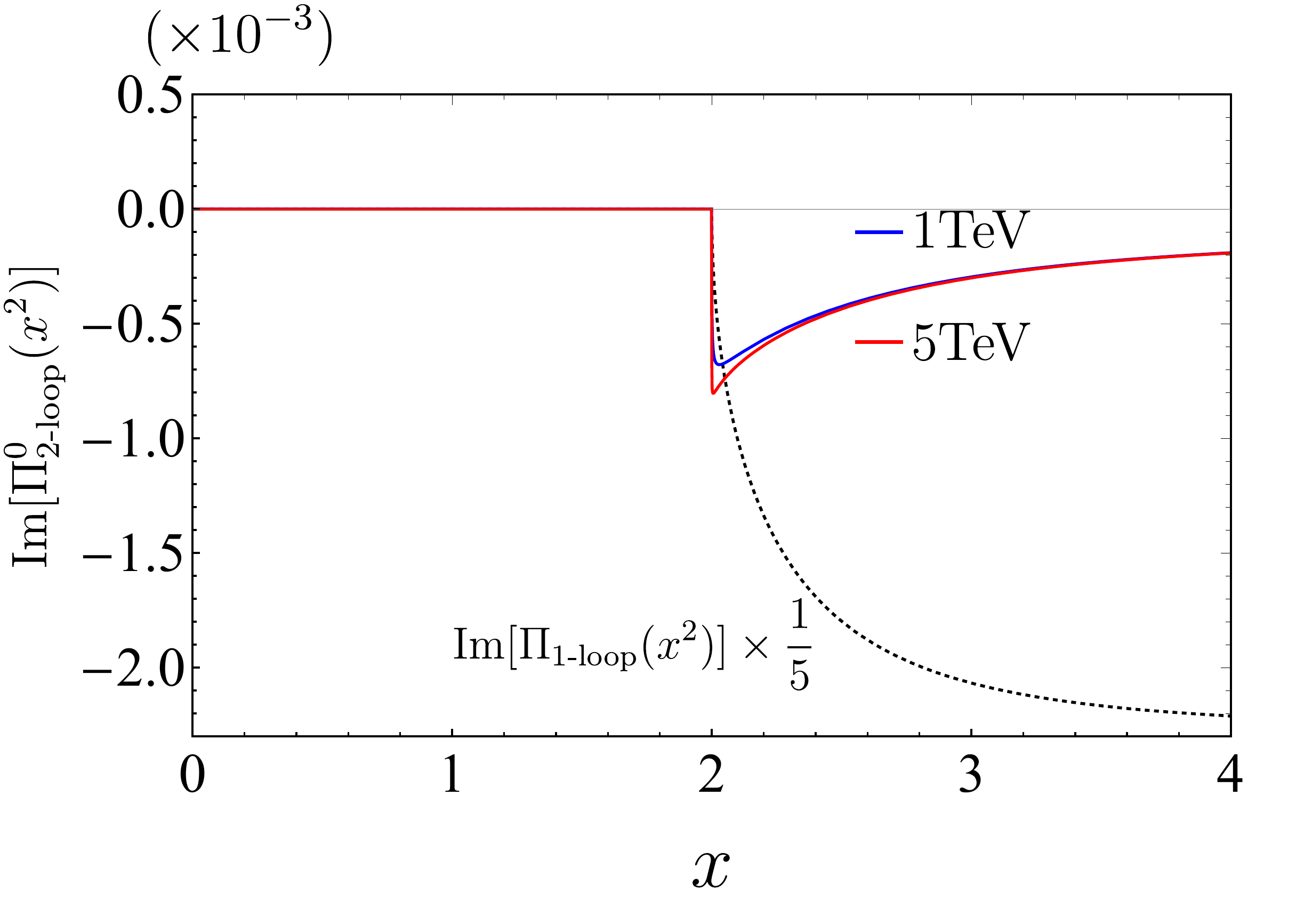}
    \\
    \includegraphics[width=82mm]{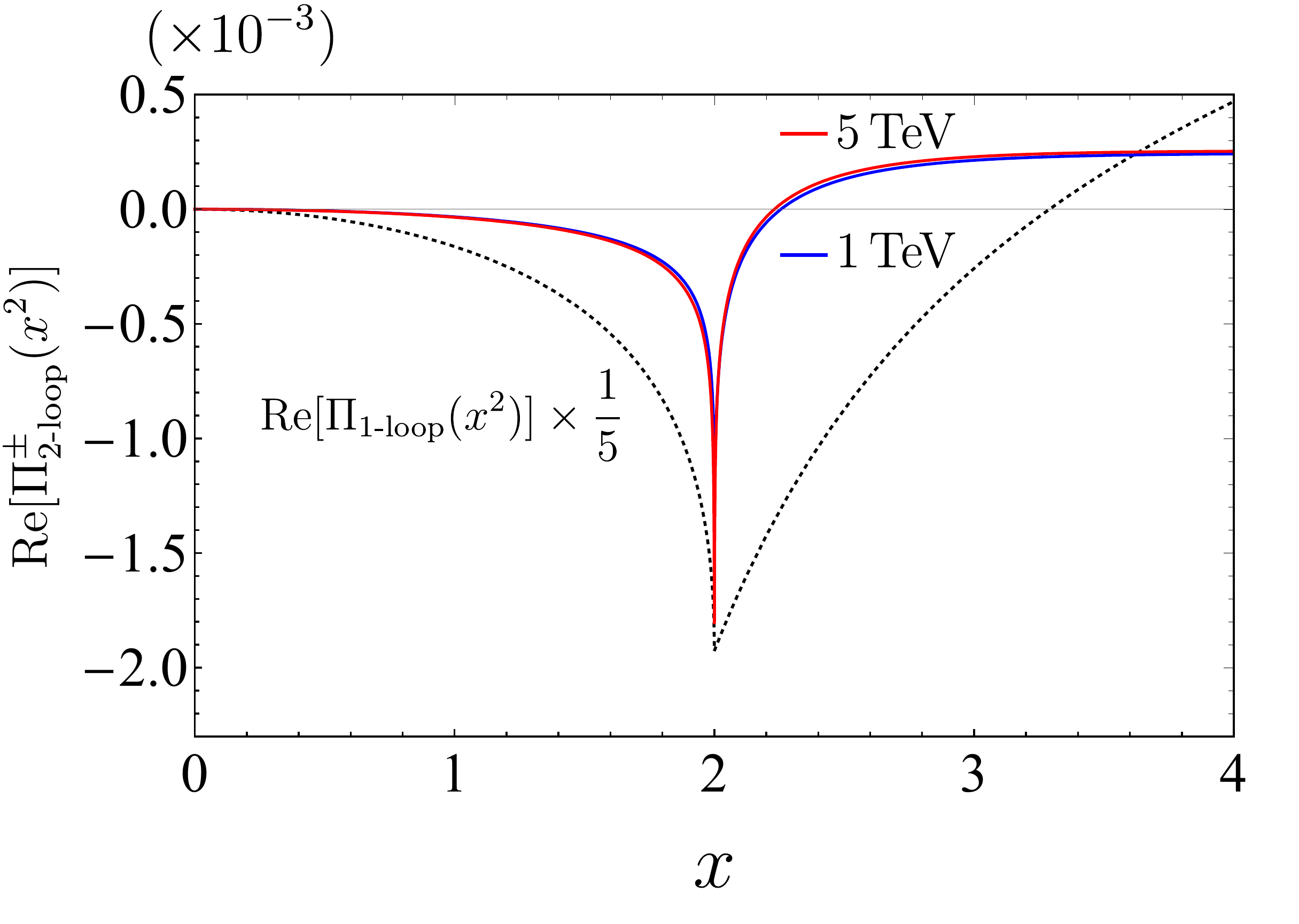}
    \qquad
    \includegraphics[width=82mm]{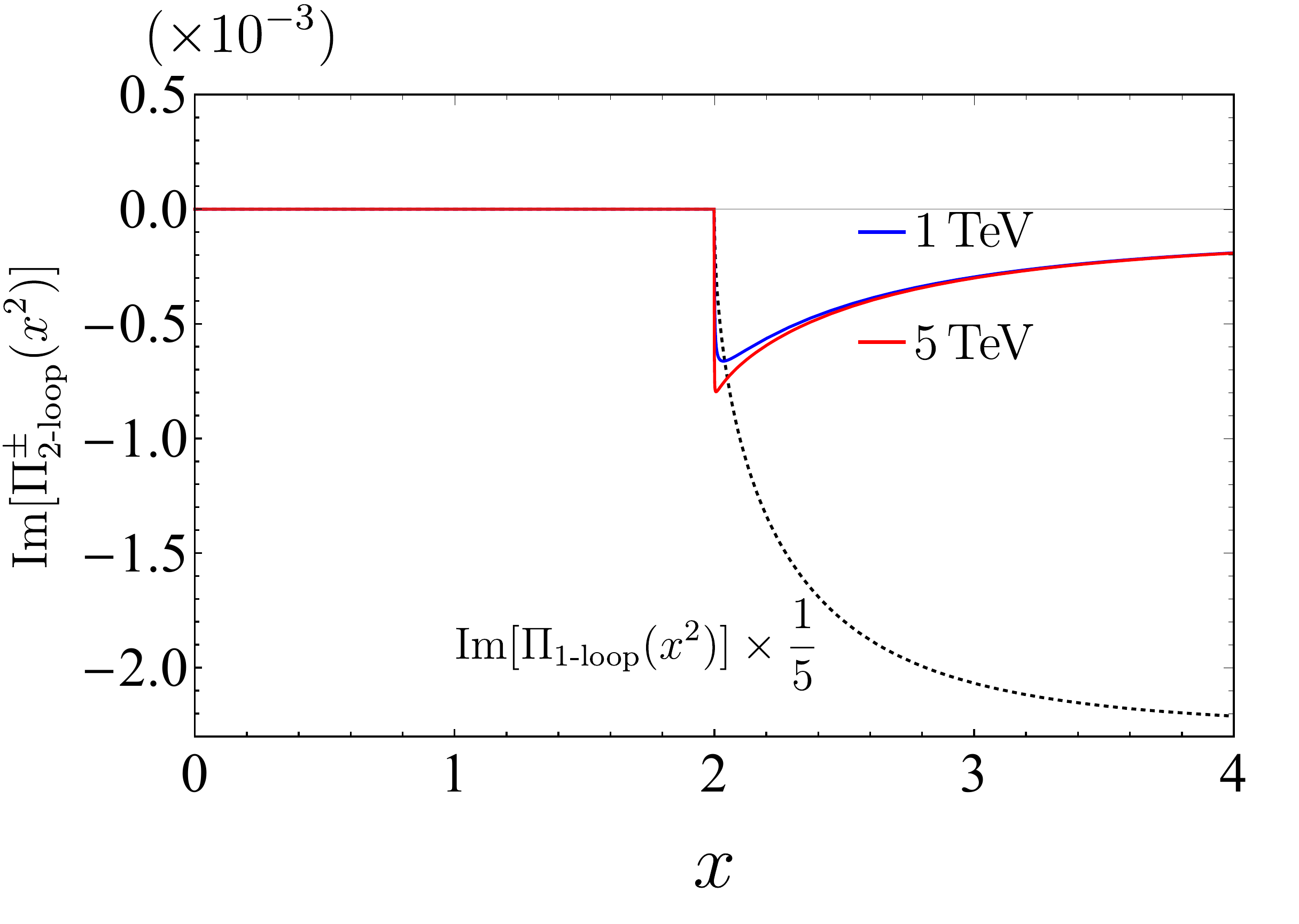}
    \caption{\small Real and imaginary parts of the next-leading perturbative correction to the self-energies of the neutral and charged electroweak gauge bosons, $\Pi_{\mathrm{2\mathchar`-loop}}^0(x^2)$ and $\Pi_{\mathrm{2\mathchar`-loop}}^\pm(x^2) \equiv \Pi_{\mathrm{2\mathchar`-loop};\,W}^\pm(x^2)$, obtained by calculating the two-loop diagrams in the text, for the cases of $m_\chi =$ 1\,TeV and 5\,TeV. The leading perturbative correction to the self-energies is also shown in each panel for comparison purposes.}
    \label{fig: 2loop}
\end{figure}

The next-leading perturbative corrections, namely $\Pi_{\mathrm{2\mathchar`-loop}}^0(x^2)$ and $\Pi_{\mathrm{2\mathchar`-loop};\,W}^\pm(x^2)$, are depicted in Fig.\,\ref{fig: 2loop} for the cases of $m_\chi =$ 1\,TeV and 5\,TeV. Real parts of the corrections have a negative bump structure at the threshold region as for the leading one, $\Pi_{\mathrm{1\mathchar`-loop}}$, shown in Fig.\,\ref{fig: 1loop}. In particular, the next-leading perturbative correction, $\Pi_{\mathrm{2\mathchar`-loop}}^0$, has a singular structure, as also can be seen from the analytical form of the next-leading perturbative corrections at the threshold region:
\begin{align}
    \Pi_{\mathrm{2\mathchar`-loop}}^0(x^2) & \simeq \frac{g^2}{8\pi}
    \left[
        \alpha\ln{(-2i\sqrt{x-2+i 0^+})} + \alpha_Z \ln{\left(-2i\sqrt{x-2+i 0^+} + \frac{m_Z}{m_{\chi}}\right)} + \mathrm{const.}
    \right] ,
    \label{eq:2loop_threshold_neutral} \\
    \Pi_{\mathrm{2\mathchar`-loop};\,W}^\pm(x^2) & \simeq \frac{g^2}{8\pi}
    \left[
        \alpha_W\ln{\left(-2i\sqrt{x-2+i 0^+} + \frac{m_W}{m_\chi}\right)} + \mathrm{const.}
    \right] ,
    \label{eq:2loop_threshold_charged}
\end{align} 
where $\alpha_Z = \alpha\,\cos^2{\theta_W}/\sin^2{\theta_W}$, $\alpha_W = \alpha /\sin^2{\theta_W}$, and $m_Z$, $m_W$ are the $Z$, $W$ boson masses, respectively, with $\theta_W$ being the Weinberg angle. The next-leading perturbative correction is thus seen to be unphysically enhanced due to the threshold singularity, but it is cured by non-perturbative corrections, as we will discuss in the next section using NR Lagrangian method of the EWIMP.

\section{Non-perturbative corrections}
\label{sec: non-perturbative corrections}

Due to the threshold singularity seen in the previous section, non-perturbative corrections must be taken into account to properly calculate the EWIMP oblique correction. Such a non-perturbative effect, which is sometimes called the Sommerfeld effect\,\cite{Hisano:2003ec, Hisano:2004ds}, originates in the attractive force caused by exchanging electroweak gauge bosons between two EWIMPs. The effect becomes sizable when the EWIMP becomes non-relativistic and causes, for instance, EWIMP bound states. Taking the non-perturbative effect into account is equivalent to resumming following ladder diagrams:
\begin{figure}[H] 
    \centering
    \includegraphics[width=84.7mm]{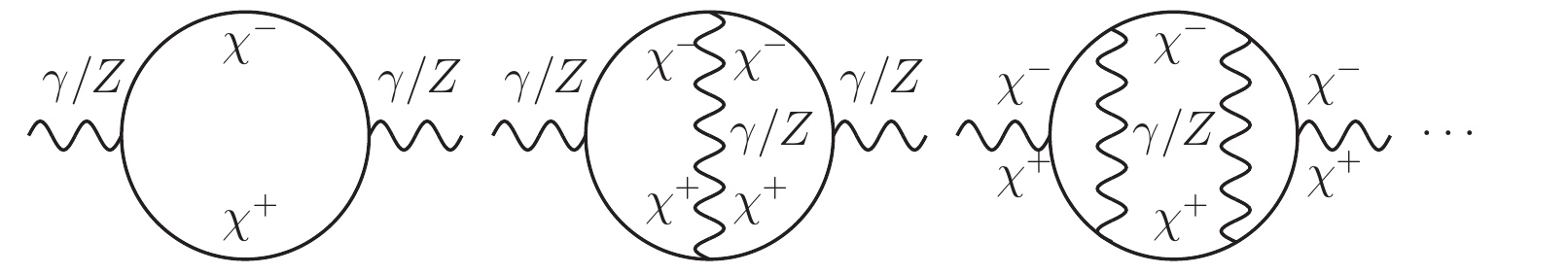}
    \qquad
    \includegraphics[width=84.7mm]{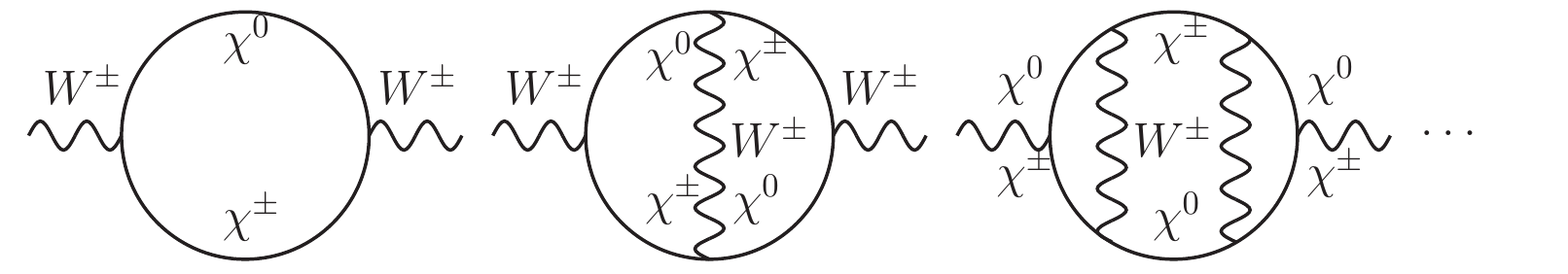}
\end{figure}
\noindent
On the other hand, instead of resumming the diagrams explicitly, the non-perturbative effect can be evaluated using the non-relativistic (NR) Lagrangian of the EWIMP, where the effect of the attractive force is obtained by solving the Schr\"{o}dinger equation for the two-body state of the EWIMPs. 

The NR Lagrangian describes the two body states which are composed of the neutral and/or charged components of the EWIMP, and it is derived from the original Lagrangian in eq.\,(\ref{eq: Lagrangian}) as follows:
\begin{align}
    \mathcal{L}_{\mathrm{2\mathchar`-body}}
    &= \int {d^3} r \, \phi^{i \dagger}_0(\Vec{r}, x)
    \left[ i\partial_0 + \frac{\nabla^2_x}{4m_{\chi}} + \frac{\nabla^2_r}{m_{\chi}} + \frac{\alpha + \alpha_Z\, e^{-m_Z |\vec{r}|}}{|\vec{r}|} \right]
    \phi^i_0(\Vec{r},x)
    \\
    &+ \int {d^3} r \, \phi^{i \dagger}_\pm(\Vec{r}, x)
    \left[ i\partial_0 + \frac{\nabla^2_x}{4m_\chi} + \frac{\nabla^2_r}{m_\chi} + \frac{\alpha_W\,e^{-m_W|\vec{r}|}}{|\vec{r}|} \right]
    \phi^i_\pm(\Vec{r},x)
    \nonumber \\
    &-g \left[e^{2 i m_{\chi} x^0}\phi^{i\dagger}_0(\Vec{0},x)\,W^3_i(x)
    + e^{2 i m_\chi x^0}\phi^{i\dagger}_+(\Vec{0},x)\,W^+_i(x)
    + e^{2 i m_\chi x^0}\phi^{i\dagger}_-(\Vec{0},x)\,W^-_i(x) + \mathrm{h.c.} \right] ,
    \nonumber
\end{align}
where $\phi^i_a(\vec{r},x)$ is the field describing the two-body state composed of $\chi^+$, $\chi^-$ ($a = 0$), and $\chi^0$, $\chi^\pm$ ($a = \pm$), with $\vec{r}$ and $x$ being the (spatially) relative and barycentric coordinates of the constituent particles, respectively, while the superscript $i = 1, 2, 3$ is the spin index of the fields. Using the above NR Lagrangian, the oblique correction to the electroweak gauge bosons is obtained as
\begin{align}
    & \Pi_{W W}(q^2) = \Pi^\pm_{\mathrm{NR}}(q^2/m_\chi^2)
    \equiv g^2\,G^\pm(\vec{0},\vec{0};q^2/m_\chi^2)/(2m_\chi^2) + Z^\pm_\mathrm{NR} , \\
    & \Pi_{\gamma \gamma}(q^2) = \sin^2\theta_W\,\Pi^0_{\mathrm{NR}}(q^2/m_{\chi}^2)
    \equiv \sin^2\theta_W\,[\, g^2\,G^0(\vec{0},\vec{0};q^2/m_{\chi}^2)/(2m_{\chi}^2) + Z^0_\mathrm{NR} \,] , \\
    & \Pi_{\gamma Z}(q^2) = \sin\theta_W \cos\theta_W\,\Pi^0_{\mathrm{NR}}(q^2/m_{\chi}^2) , \\
    & \Pi_{Z Z}(q^2) = \cos^2\theta_W\,\Pi^0_{\mathrm{NR}}(q^2/m_{\chi}^2) ,
    \label{eq: NR oblique correction}
\end{align}
where $\Pi^0_\mathrm{NR}(x^2)$ and $\Pi^\pm_\mathrm{NR}(x^2)$, or to be more precise, Green functions $G^0(\vec{0},\vec{0};\,x^2)$ and $G^\pm(\vec{0},\vec{0};\,x^2)$ are obtained by solving the following Schr\"{o}dinger equations derived by the NR Lagrangian:
\begin{align}
    & \left[\frac{\nabla^2_r}{m_{\chi}} + \frac{\alpha + \alpha_Z\, e^{-m_Z |\vec{r}|}}{|\vec{r}|} + (x - 2)\,m_{\chi} \right] G^0(\vec{r},\vec{r}\,'; x^2)
    = \delta(\vec{r}-\vec{r}\,') , \\
    & \left[\frac{\nabla^2_r}{m_\chi} + \frac{\alpha_W\,e^{-m_W|\vec{r}|}}{|\vec{r}|} + (x - 2)\,m_\chi \right] G^\pm(\vec{r},\vec{r}\,'; x^2)
    = \delta(\vec{r}-\vec{r}\,') .
\end{align}
On the other hand, $Z_\mathrm{NR}^0$ and $Z_\mathrm{NR}^\pm$ appearing in the functions $\Pi^0_\mathrm{NR}(x^2)$ and $\Pi^\pm_\mathrm{NR}(x^2)$ are constants that do not depend on $q^2$, and those are related to the renormalization for kinetic terms (two-point functions) of the electroweak gauge bosons. Here, it is worth emphasizing that the oblique correction evaluated by the NR Lagrangian is valid at the threshold region, namely $|x - 2| \ll 1$, while the perturbative calculation of the correction is reliable at $|x - 2| \gg \alpha_W^2 \sim 10^{-3}$ due to the threshold singularity. Moreover, the NR correction is significantly affected by EWIMP bound states at $x \leq 2$. We therefore fix the constants $Z_\mathrm{NR}^0$ and $Z_\mathrm{NR}^\pm$ using the following matching conditions,
\begin{align}
    \Pi_{\mathrm{NR}}^{0/\pm}(x^2) = \Pi_{\mathrm{1\mathchar`-loop}}(x^2) + \Pi_{\mathrm{2\mathchar`-loop}}^{0/\pm}(x^2)
    \qquad {\rm at} ~~ x = 2.01 ,
\end{align}
and use the result of the NR Lagrangian at $x \leq 2.01$ in order to evaluate the oblique correction, while use that of the perturbative calculation at $x \geq 2.01$ in the region above the threshold, $x \geq 2$. In the region below the threshold, $x \leq 2$, on the other hand, we use the result of perturbative calculation at $x \leq x_0$, with $x_0$ being the point where both the results of the NR Lagrangian and perturbative calculations coincide at the region below that the lowest-energy bound state (lowest-energy resonance) is located. We use the result of the NR Lagrangian at $x_0 \leq x \leq 2$.\footnote{We have confirmed numerically that $x_0$ is uniquely determined, namely the procedure to switch the result of the NR Lagrangian and that of the perturbative calculation at $x = x_0$ works well for all $m_{\chi^\pm}$ and $m_\chi$ of interests.} 

\begin{figure}[t] 
    \centering
    \includegraphics[width=82mm]{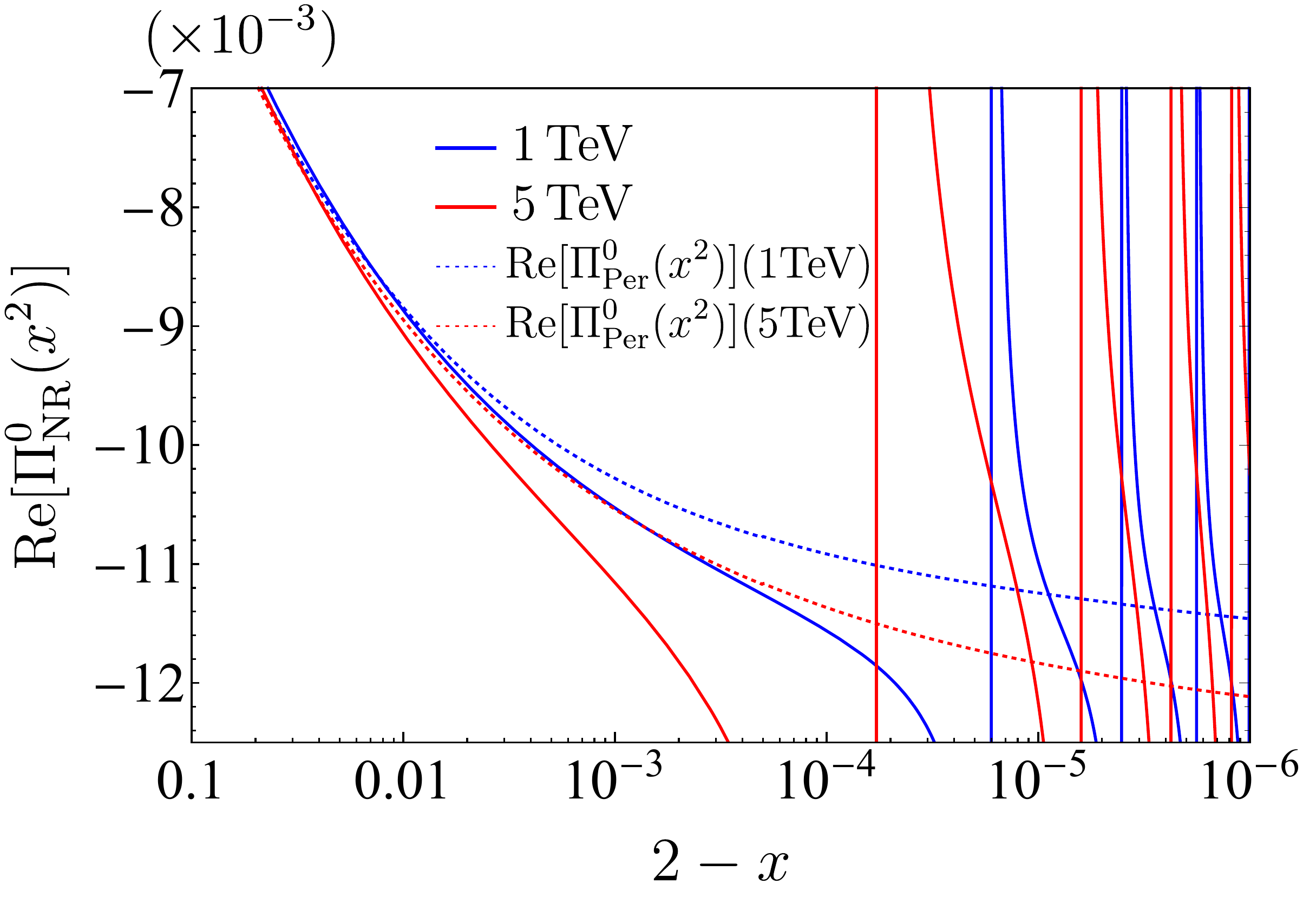}
    \qquad
    \includegraphics[width=82mm]{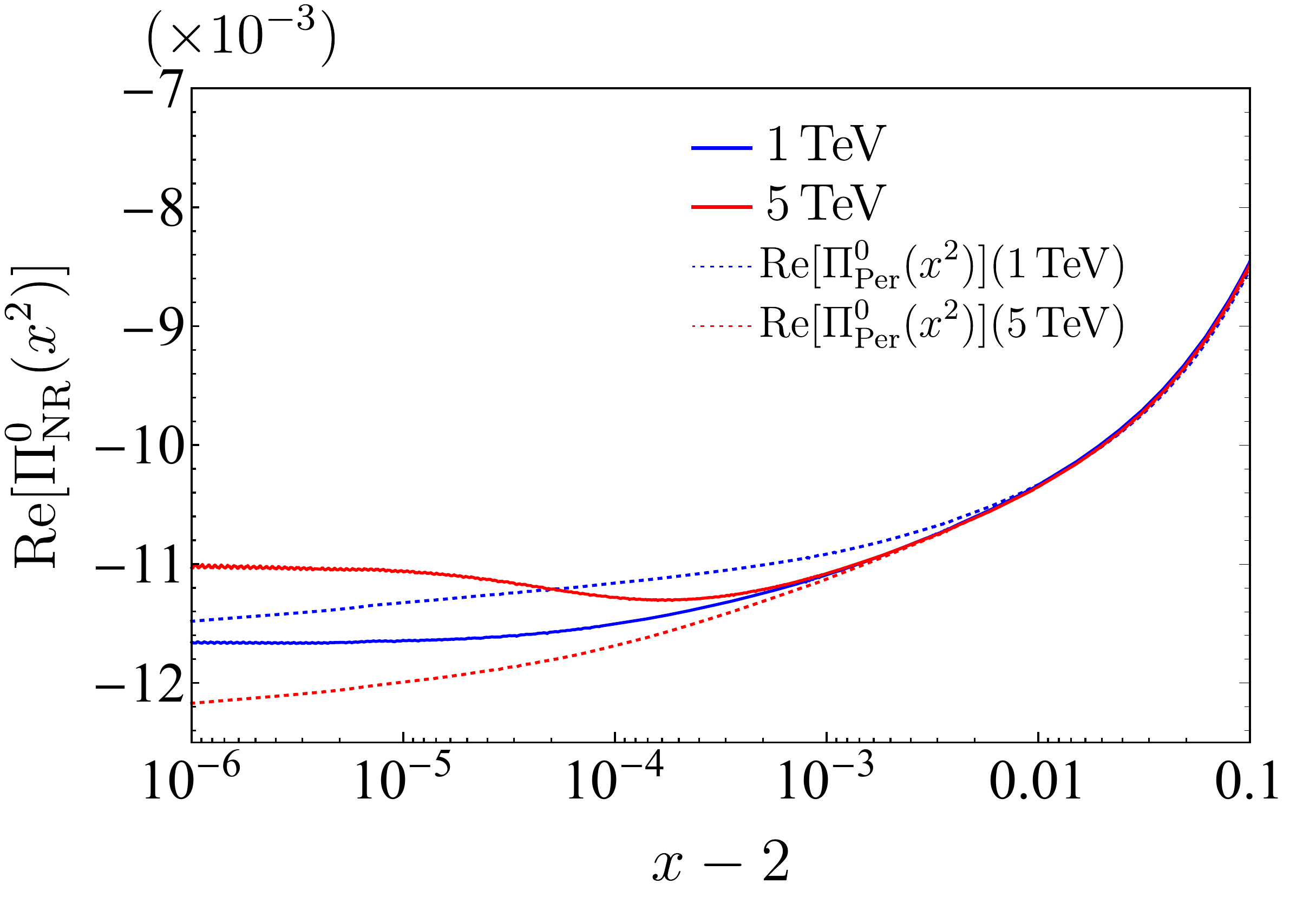}
    \\
    \includegraphics[width=82mm]{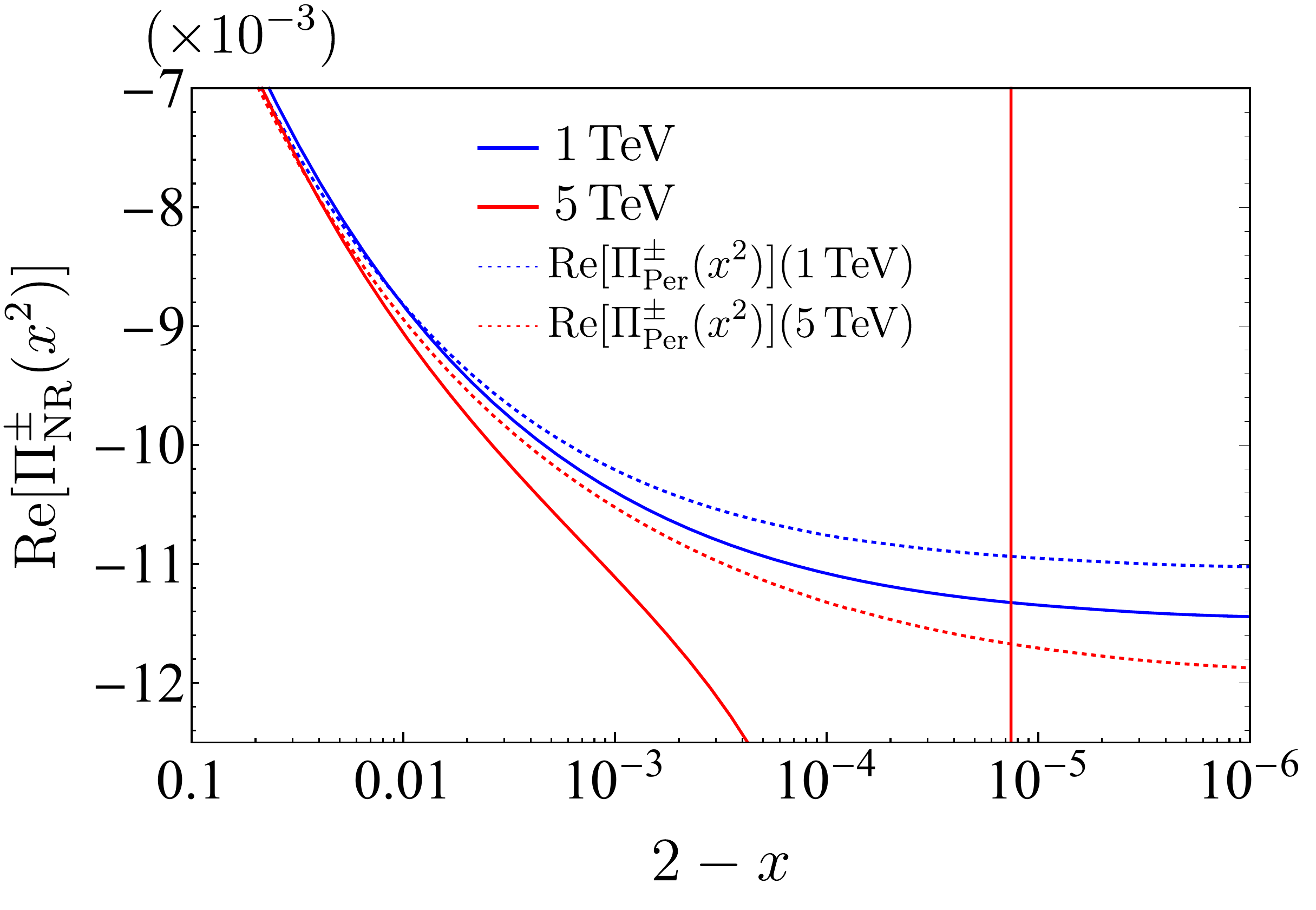}
    \qquad
    \includegraphics[width=82mm]{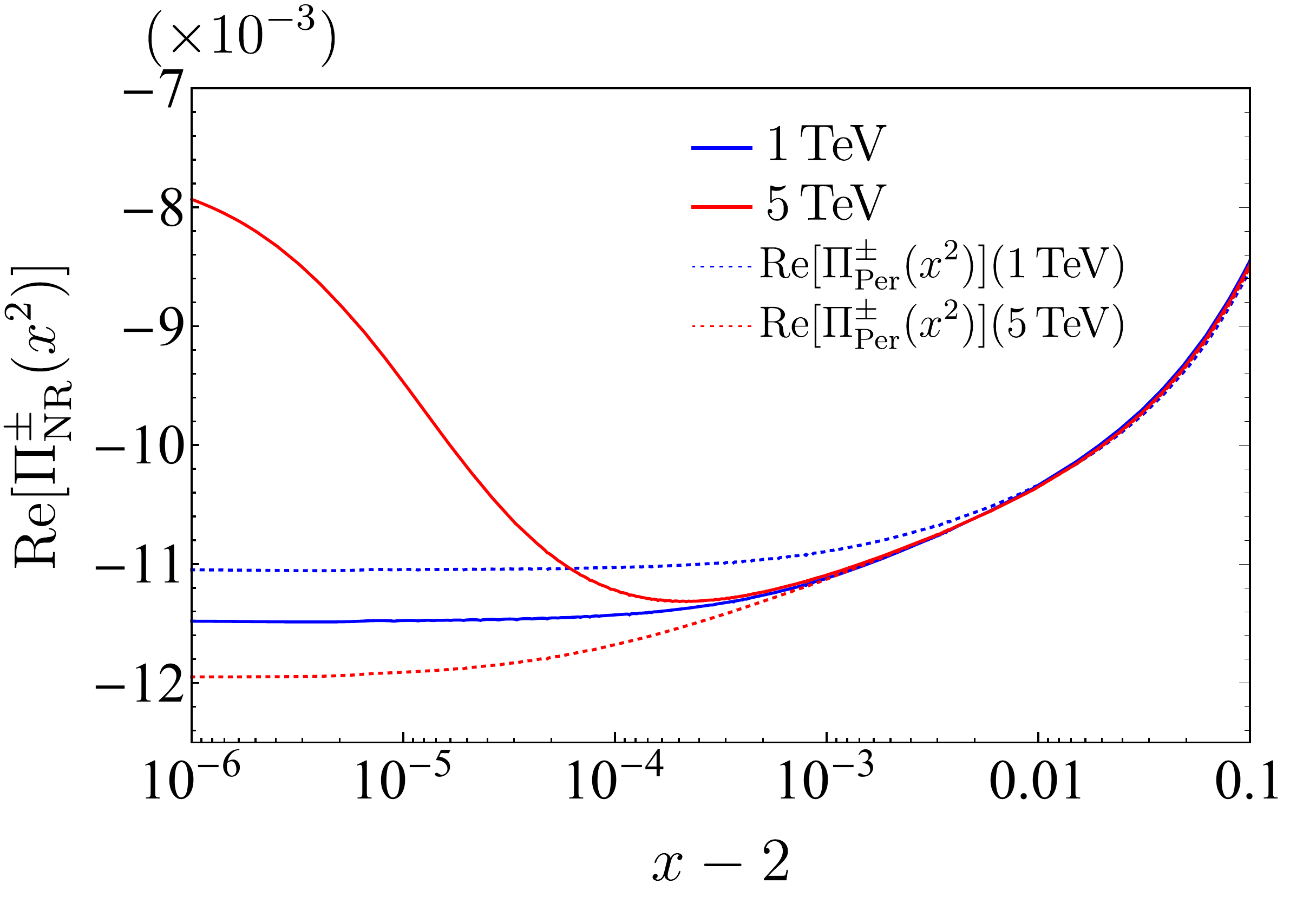}
    \caption{\small Real parts of the non-perturbative corrections to the self-energies of the neutral and charged electroweak gauge bosons, $\Pi_{\mathrm{NR}}^0(x^2)$ and $\Pi_{\mathrm{NR}}^\pm(x^2)$, are depicted for the cases of $m_{\chi^\pm}$, $m_\chi =$ 1\,TeV and 5\,TeV. For comparison purposes, real parts of the perturbative corrections, $\Pi_{\mathrm{Per}}^0(x^2) \equiv \Pi_{\mathrm{1\mathchar`-loop}}(x^2) + \Pi_{\mathrm{2\mathchar`-loop}}^0(x^2)$ and $\Pi_{\mathrm{Per}}^\pm(x^2) \equiv \Pi_{\mathrm{1\mathchar`-loop}}(x^2) + \Pi_{\mathrm{2\mathchar`-loop}}^\pm(x^2)$, to the self-energies are also shown in each panel.}
    \label{fig: NR}
\end{figure}

\begin{figure}[t]
    \centering
    \includegraphics[width=82mm]{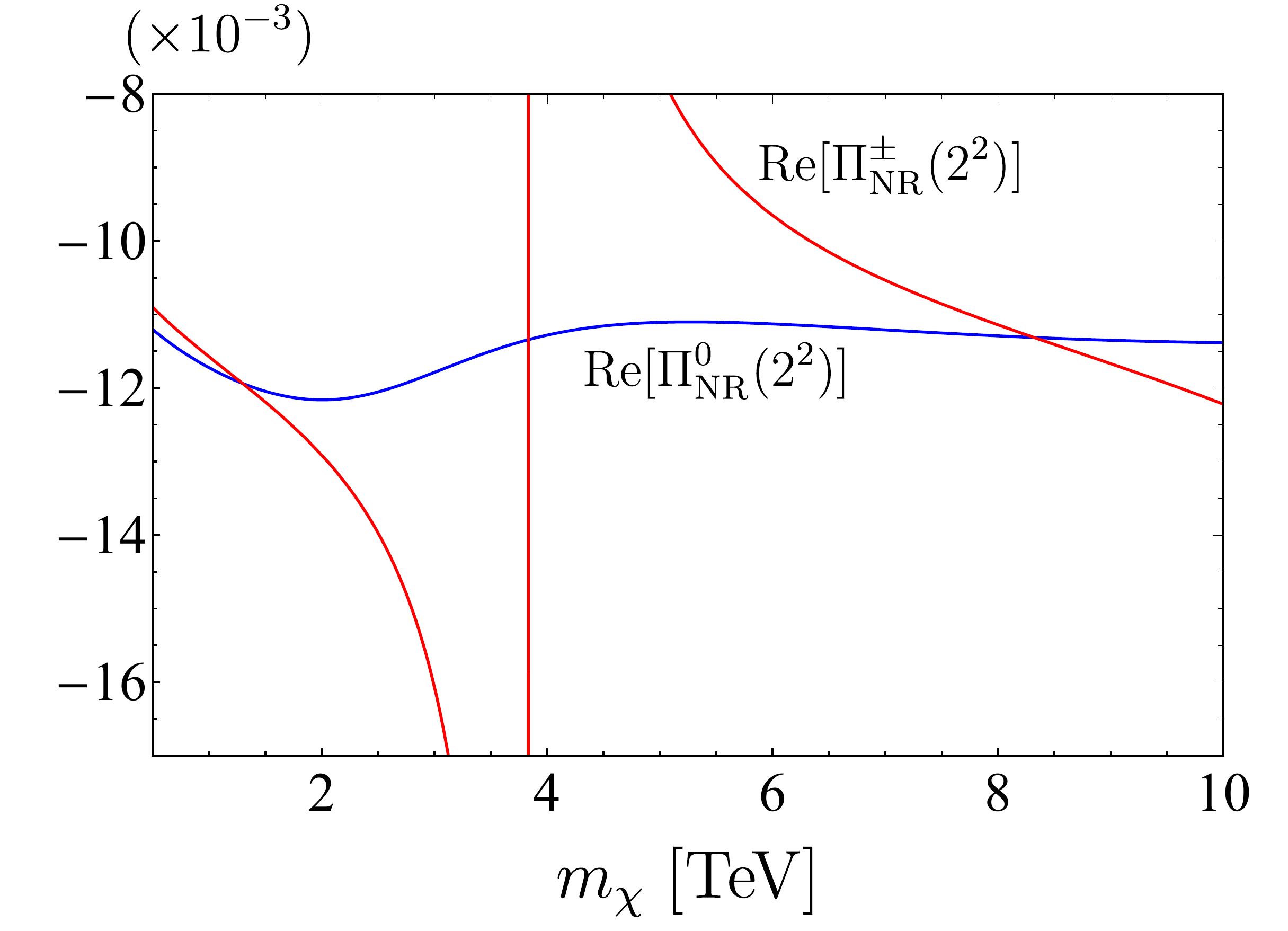}
    \caption{\small Values of the corrections ${\rm Re}[\Pi^0_{\mathrm{NR}}(2^2)]$ and ${\rm Re}[\Pi^\pm_{\mathrm{NR}}(2^2)]$ as a function of $m_\chi$ and $m_{\chi^\pm}$, respectively.}
    \label{fig: NR threshold value}
\end{figure}

Real parts of the non-perturbative corrections, ${\rm Re}[\Pi_{\mathrm{NR}}^0(x^2)]$ and ${\rm Re}[\Pi_{\mathrm{NR}}^\pm(x^2)]$, at the threshold region are shown in Fig.\,\ref{fig: NR}. As seen in the right panels, the corrections, in particular the one for the neutral electroweak gauge bosons ${\rm Re}[\Pi_{\mathrm{NR}}^0(x^2)]$, are not seen to be diverged anymore,\footnote{We discuss the behavior of the function $\Pi^0_{\mathrm{NR}}(x^2)$ at the threshold region by taking the limit $x \to 2$ from above, because the function is not continuous at $x \leq 2$ due to the divergence caused by infinitely many bound states.} unlike the perturbative correction ${\rm Re}[\Pi^0_{\mathrm{2\mathchar`-loop}}(x^2)]$ discussed in the previous section. In order to see this fact more explicitly, we show the values of ${\rm Re}[\Pi^0_{\mathrm{NR}}(2^2)]$ and ${\rm Re}[\Pi^\pm_{\mathrm{NR}}(2^2)]$ in Fig.\,\ref{fig: NR threshold value} as functions of $m_{\chi}$. It is seen that the value of ${\rm Re}[\Pi^0_{\mathrm{NR}}(2^2)]$ never diverges at any mass of $m_{\chi^\pm}$. On the other hand, the value of ${\rm Re}[\Pi^\pm_{\mathrm{NR}}(2^2)]$ diverges at the specific mass of $m_\chi$, namely $m_\chi \sim 4$\,TeV. This is due to the so-called zero-energy resonance, where the binding energy of the bound state becomes zero at this specific mass because of the nature of the Yukawa potential. The bound state is thus located exactly on the threshold and it makes the value of ${\rm Re}[\Pi^\pm_{\mathrm{NR}}(2^2)]$ diverged.

It is seen in the left panels of Fig.\,\ref{fig: NR} that the non-perturbative corrections, ${\rm Re}[\Pi_{\mathrm{NR}}^0(x^2)]$ and ${\rm Re}[\Pi_{\mathrm{NR}}^\pm(x^2)]$, are significantly different from the perturbative ones, ${\rm Re}[\Pi_{\mathrm{Per}}^0(x^2)]$ and ${\rm Re}[\Pi_{\mathrm{Per}}^\pm(x^2)]$, at $x \leq 2$. This is due to EWIMP bound states caused by the electroweak interaction. With $E^{0/\pm}_n$ being the binding energy of the $n$-th bound state in $\Pi_{\mathrm{NR}}^{0/\pm}(x^2)$, the correction can be written as
\begin{align}
    \Pi^{0/\pm}_{\mathrm{NR}}(x^2)
    = \Pi_\mathrm{Reg}^{0/\pm}(x^2)
    + \sum_n \frac{(g^{0/\pm}_n)^2}{x^2 - (x^{0/\pm}_n)^2 + 2 i \gamma^{0/\pm}_n} ,
    \label{eq: dispersion}
\end{align}
where $\Pi^{0/\pm}_\mathrm{Reg}(x^2)$ is the non-singular part of the correction that is obtained by subtracting singular parts (poles representing the bound states) from the correction, while $x^0_n \equiv 2 - E^0_n/m_{\chi}$, $x^\pm_n \equiv 2 - E^\pm_n/m_\chi$, and $\gamma^0_n \equiv \Gamma^0_n/m_{\chi}$, $\gamma^\pm_n \equiv \Gamma^\pm_n/m_\chi$ with $\Gamma^0_n$ and $\Gamma^\pm_n$ being decay widths of $n$-th bound states appearing in the functions $\Pi_{\mathrm{NR}}^0(x^2)$ and $\Pi_{\mathrm{NR}}^\pm(x^2)$, respectively. Residues of the poles are depicted by $(g^{0/\pm}_n)^2$, where $g^{0/\pm}_n$ represent couplings (strength of interactions) between the bound states and the electroweak gauge bosons. As a result, the non-perturbative corrections, ${\rm Re}[\Pi_{\mathrm{NR}}^0(x^2)]$ and ${\rm Re}[\Pi_{\mathrm{NR}}^\pm(x^2)]$, behaves at $x \leq 2$ as those seen in the left panels of Fig.\,\ref{fig: NR}. Here, it is worth notifying that infinitely many bound states contribute to $\Pi^0_{\mathrm{NR}}(x^2)$ due to the nature of Coulomb potential, while a finite number of bound states contributes to $\Pi^\pm_{\mathrm{NR}}(x^2)$ because only Yukawa potential exists in the two-body system composed of $\chi^0$ and $\chi^\pm$. In fact, the number of bound states is zero when $m_\chi = 1$\,TeV, while it becomes one when $m_\chi = 5$\,TeV, as seen in the bottom-left panel of Fig.\,\ref{fig: NR}.

\begin{figure}[t] 
    \centering
    \includegraphics[width=82mm]{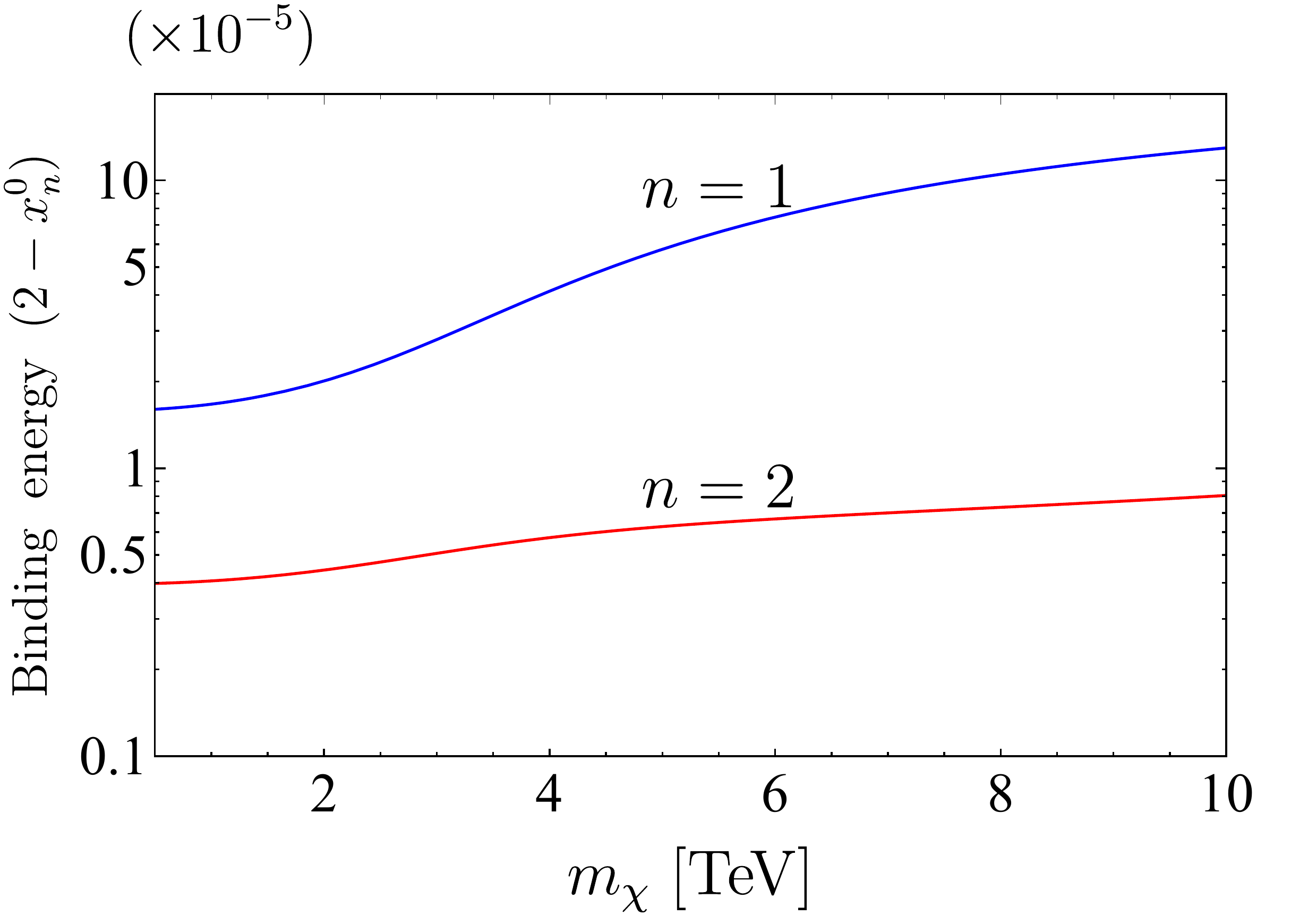}
    \qquad
    \includegraphics[width=82mm]{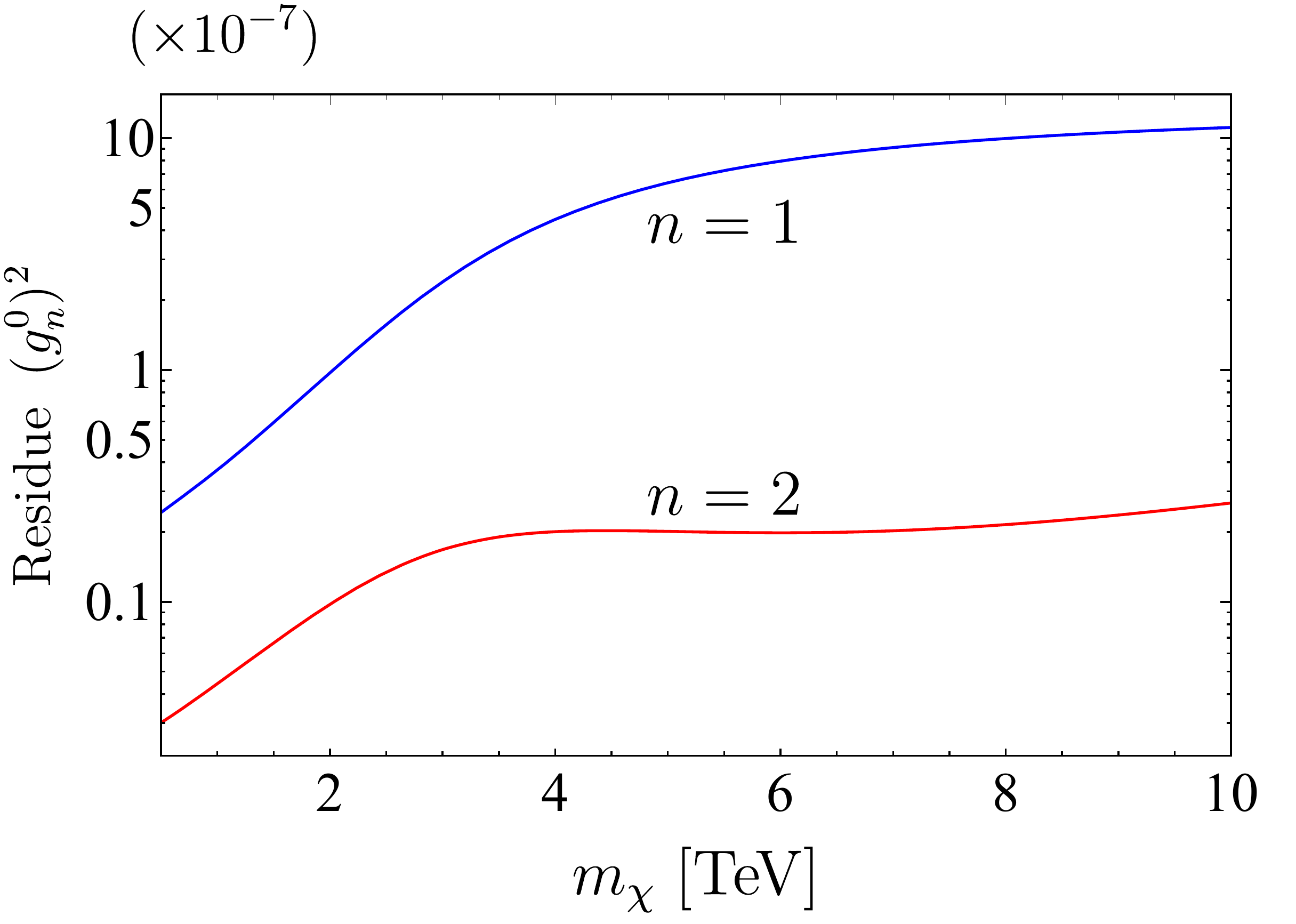}
    \\
    \includegraphics[width=82mm]{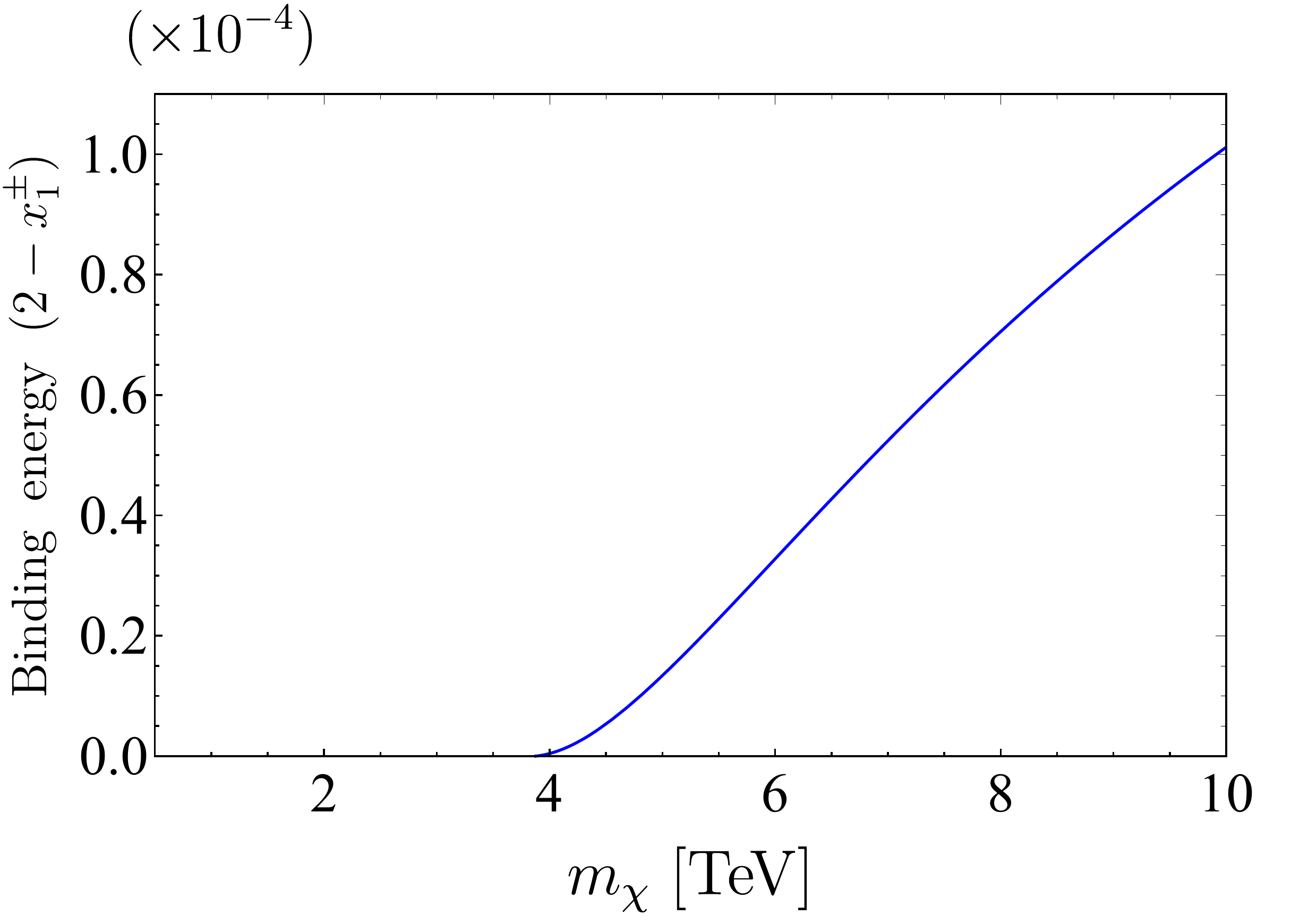}
    \qquad
    \includegraphics[width=82mm]{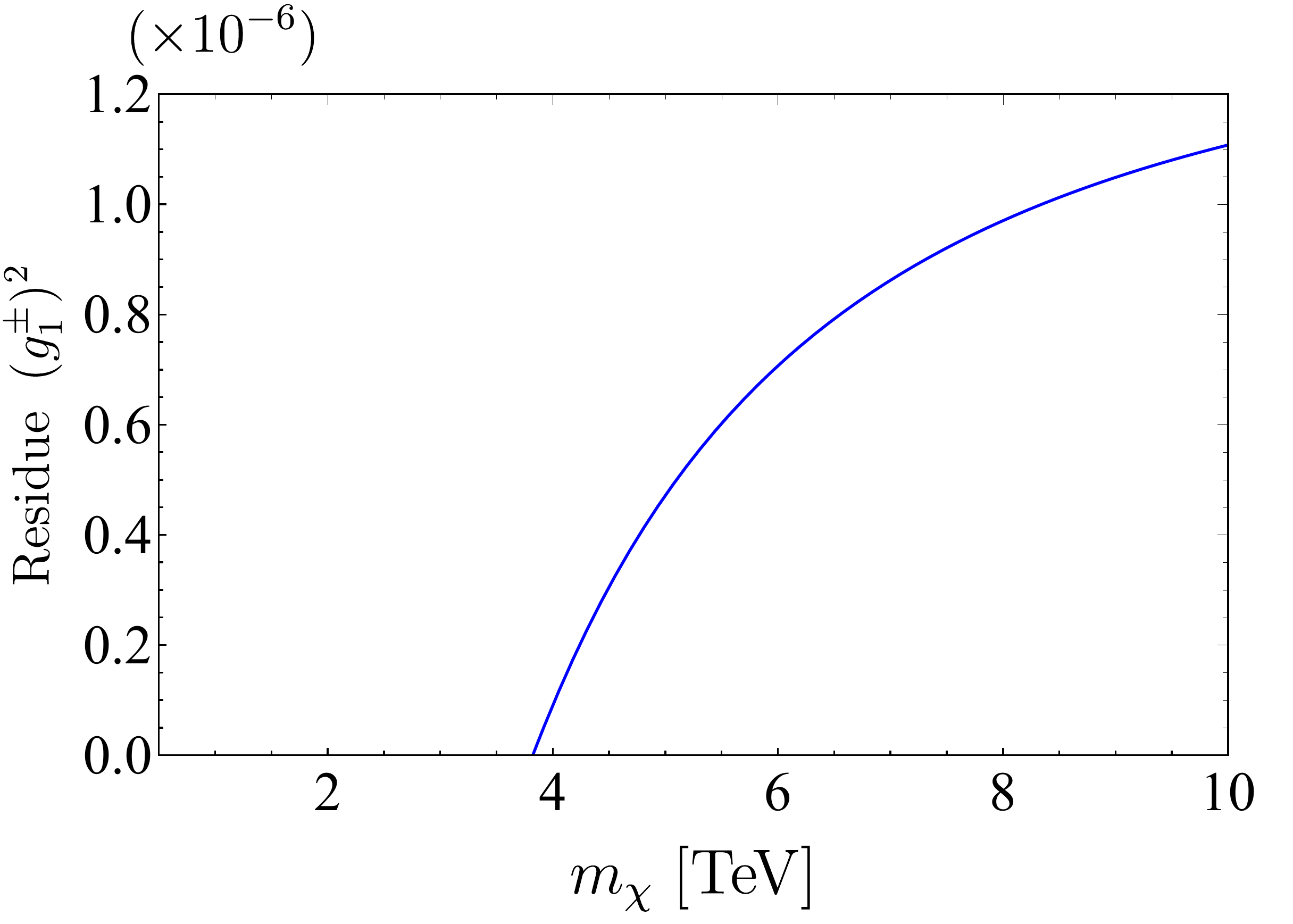}
    \caption{\small Binding energies $2 - x^{0/\pm}_n$ (left panels) and residues $g^{0/\pm}_n$ (right panels) are depicted for $n = 1, 2$ in the case of $\Pi^0_\mathrm{NR}(x^2)$ and $n = 1$ in the case of $\Pi^\pm_\mathrm{NR}(x^2)$ as functions of  $m_\chi$.}
    \label{fig: pole structures}
\end{figure}

The values of $g^{0/\pm}_n$ and $x^{0/\pm}_n$ are obtained numerically from the function $\Pi^{0/\pm}_\mathrm{NR}(x^2)$ as shown in Fig.\,\ref{fig: pole structures}, where the binding energy $2 - x^{0/\pm}_n$ and residue $(g^{0/\pm}_n)^2$ are depicted for $\Pi^0_\mathrm{NR}(x^2)$ and $\Pi^\pm_\mathrm{NR}(x^2)$ as  functions of $m_{\chi}$. It is seen from the bottom panels that the bound state exists when $m_\chi \gtrsim 4$\,TeV, meaning that the so-called zero-energy bound state appears when $m_\chi \simeq 4$\,TeV, as we deduced in Fig.\,\ref{fig: NR threshold value}. On the other hand, as seen in the top panels, bound states always exist with irrespective to $m_{\chi}$. In fact, not only the first and second bound states shown in the panels but also infinitely many higher ones exist. Because high enough bound states are governed almost solely by the Coulomb part of the potential, their binding energies are estimated to be $2 - x^0_n  \simeq  \alpha^2/(4n^2)$. On the other hand, because the residues are written as $(g^{0/\pm}_n)^2 = 2\,g^2\, |\Psi^{0/\pm}_n(\vec{0})|^2$ in general with $\Psi^{0/\pm}_n(\vec{x})$ being the wave function (normalized by $m_\chi$) describing the $n$-th bound state, the residues of the high enough bound states can also be estimated as $(g^0_n)^2 \simeq g^2\,\alpha^3/(4\pi n^{3})$.

Finally, the decay width of the bound state denoted by $\gamma^{0/\pm}_n$ in eq.\,(\ref{eq: dispersion}) is written as follows:
\begin{align}
    \gamma^{0/\pm}_n \simeq \frac{g^2}{48\pi} (g^{0/\pm}_n)^2
    + N_F \frac{g^2}{48\pi} (g^{0/\pm}_n)^2 = \frac{25\,g^2}{48\pi} (g^{0/\pm}_n)^2 ,
\end{align}
when the EWIMP is enough heavier than SM particles. The first term is the contribution from decays into electroweak gauge bosons and Higgs boson, while the second term is from decays into SM fermions with $N_F = 6 + 3 \times 6$ being the number of the left-handed fermions (leptons and quarks). As seen in the above formula, the decay width is contributed from various annihilation processes between constituent particles of the bound states. On the other hand, there is another contribution from the decay of the constituent particle $\chi^\pm$, however it is negligibly small compared to those from the annihilation, so that we do not include such a contribution in the decay width of $\gamma^{0/\pm}_n$.

\section{EWIMP signals at hadron colliders}
\label{sec: collider signals}

We are now at the position to discuss EWIMP signals at hadron collider experiments based on the result obtained in the previous sections. The effect of the EWIMP on the Drell-Yan processes for SM lepton pair productions are already discussed in some details in section\,\ref{sec: oblique correction}. In order to evaluate realistic signals at hadron collider experiments, however, we also have to take into account the effect of the finite energy resolution for the lepton measurement. Then, the differential cross section at a certain invariant mass of the final state lepton pair is obtained by following formulae:
\begin{align}
    & \frac{d\sigma_\mathrm{SM}^{\rm Neutral/Charged}}{d m_{\ell \ell}} \propto
    \int d\delta \left|\mathcal{M}_\mathrm{LO}^{\rm Neutral/Charged}(m_{\ell \ell}+\delta) \right|^2 f(\delta,\,\sigma) , \\
    & \frac{d\sigma_\mathrm{BSM}^{\rm Neutral/Charged}}{d m_{\ell \ell}} \propto
    \int d\delta \left|\mathcal{M}_\mathrm{LO}^{\rm Neutral/Charged}(m_{\ell \ell}+\delta) +
    \mathcal{M}_\mathrm{EWIMP}^{\rm Neutral/Charged}(m_{\ell \ell}+\delta) \right|^2 f(\delta,\,\sigma) ,
\end{align}
where $f(\delta,\,\sigma)$ is the smearing function representing the finite energy resolution with $\delta$ and $\sigma$ being the fluctuation around $m_{\ell \ell}$ and the size of the energy resolution, respectively. We adopt Gauss distribution function for $f(\delta,\,\sigma) = \exp [-\delta^2/(2 \sigma^2) ]/(2 \pi \sigma^2)^{1/2}$. Since the energy resolution of the lepton measurement is currently comparable to or less than 1\%\,\cite{Adzic:2007mi}, we set $\sigma$ to be $0.1\,\%$, namely $\sigma = 10^{-3}\,m_{\ell \ell}$, as a optimistic expectation for future hadron (and high-energy lepton) colliders.

\begin{figure}[h!] 
    \centering
    \includegraphics[width=82mm]{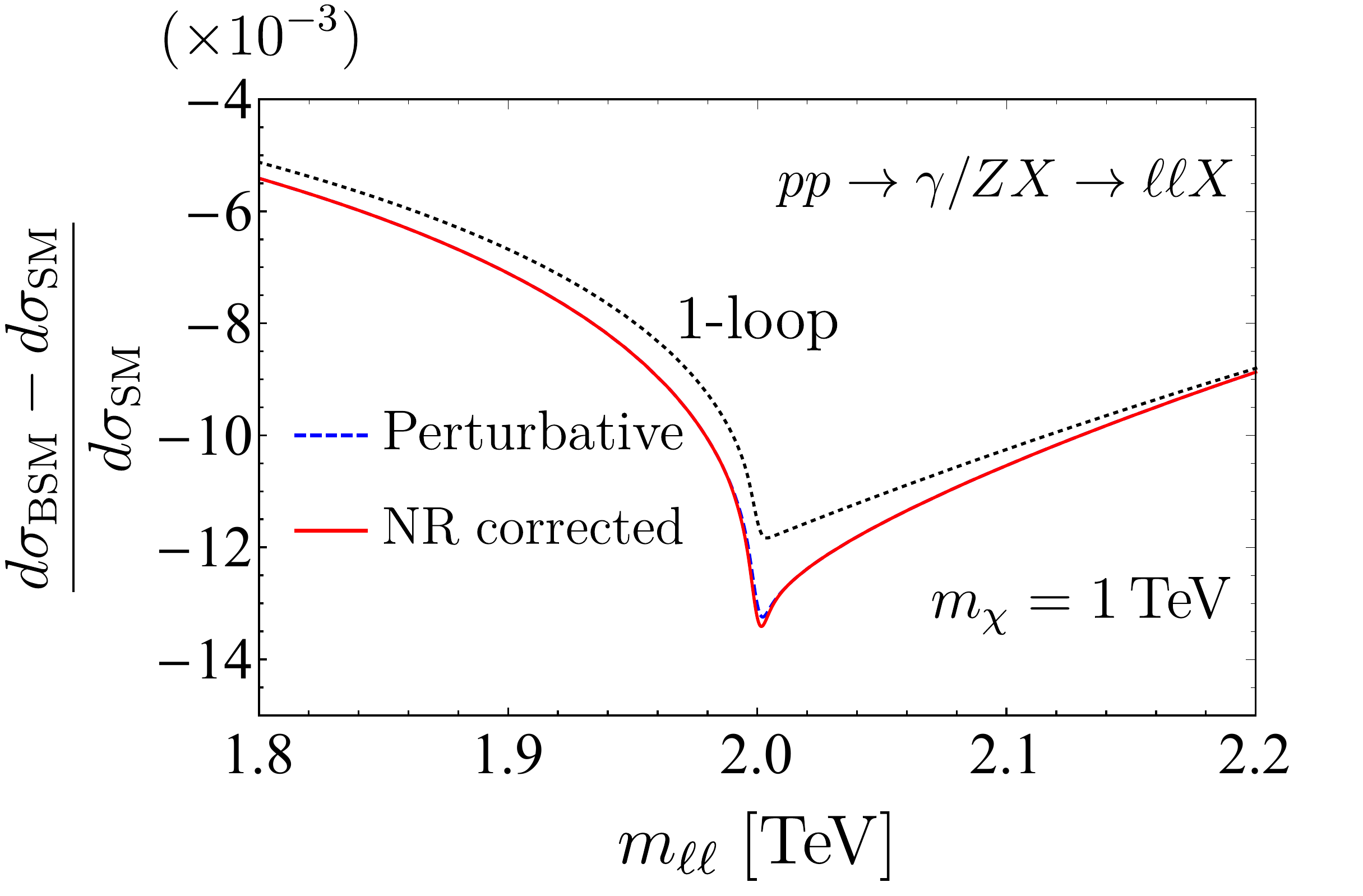}
    \qquad
    \includegraphics[width=82mm]{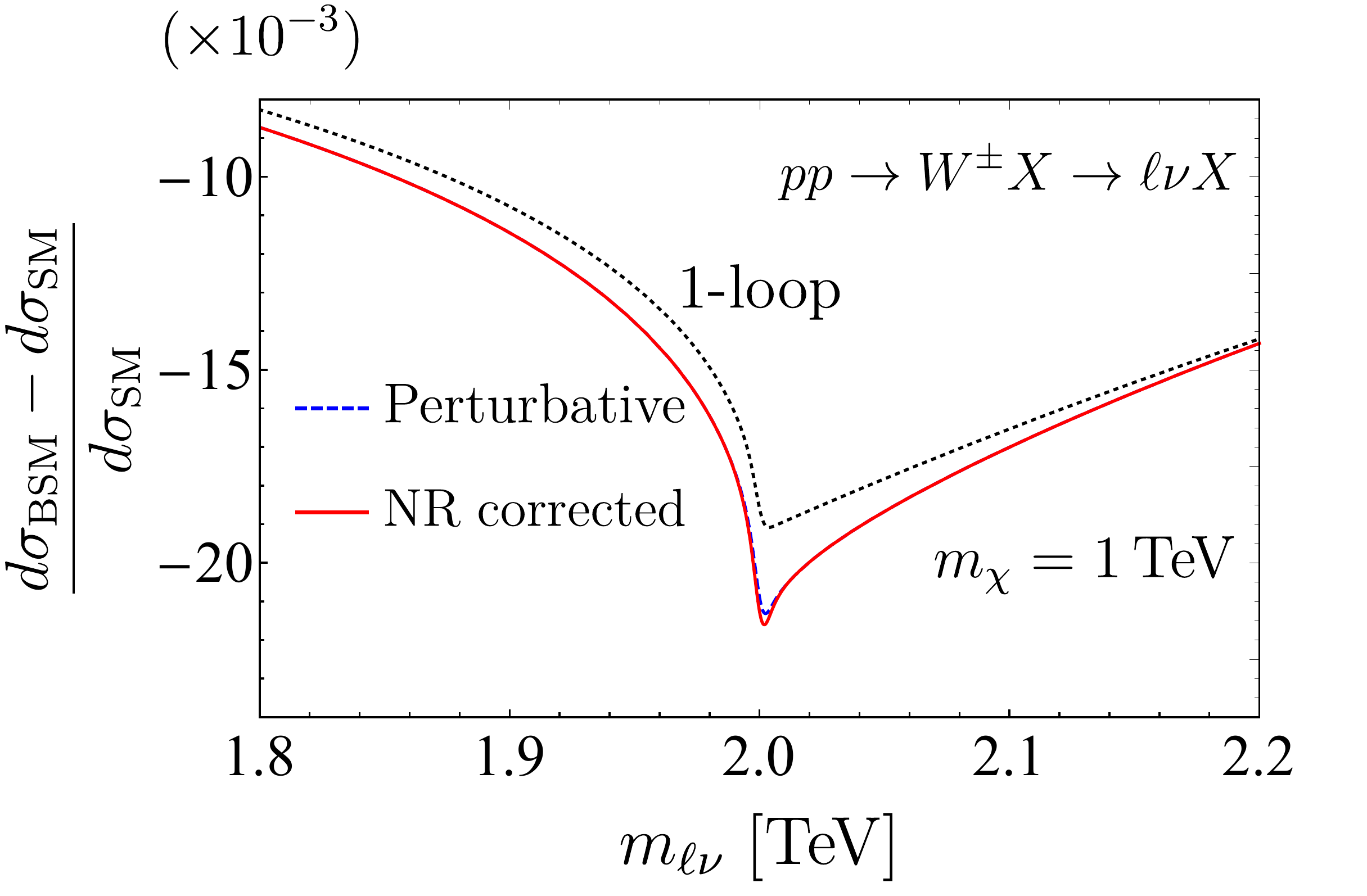}
    \\
    \includegraphics[width=82mm]{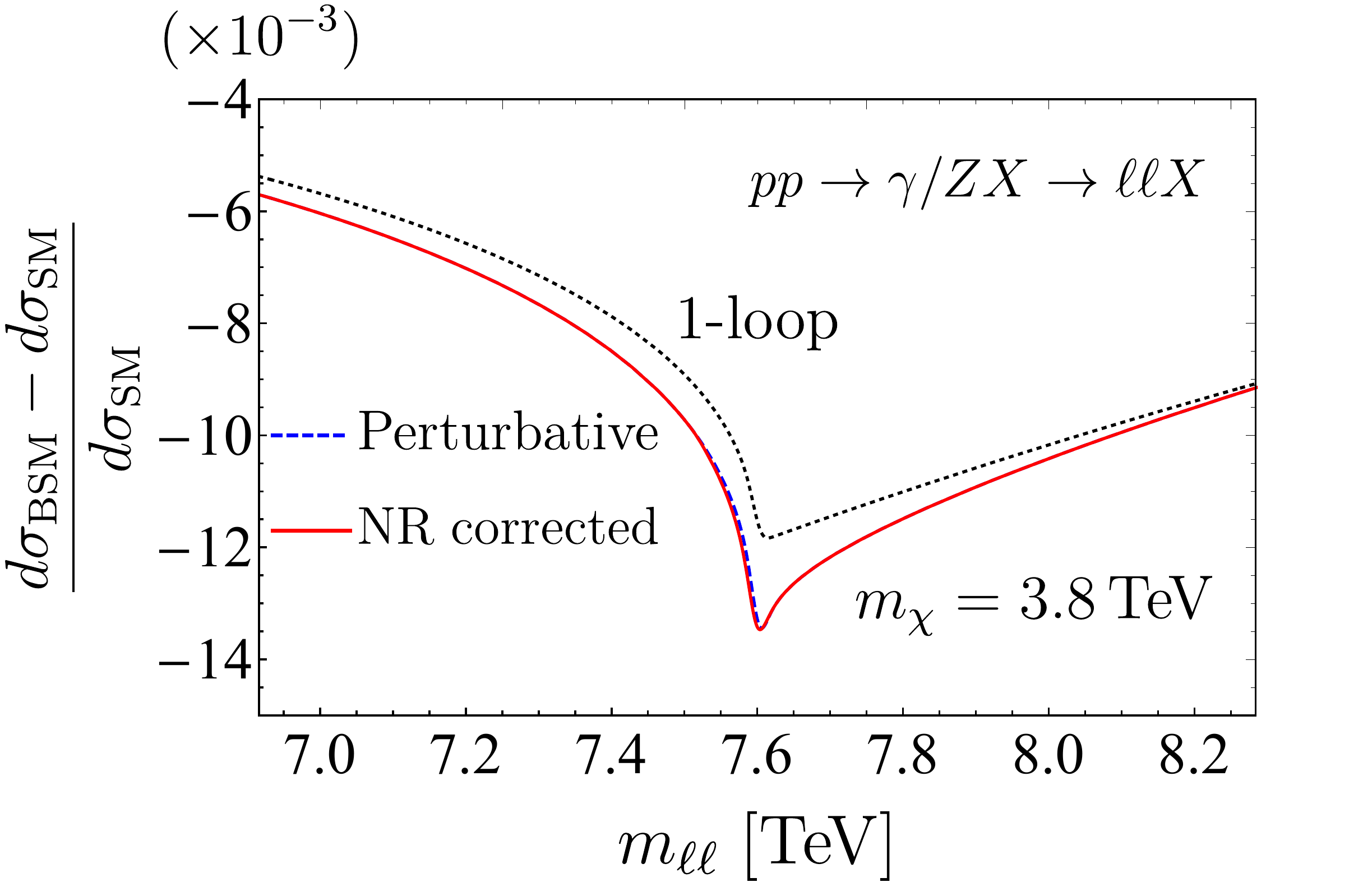}
    \qquad
    \includegraphics[width=82mm]{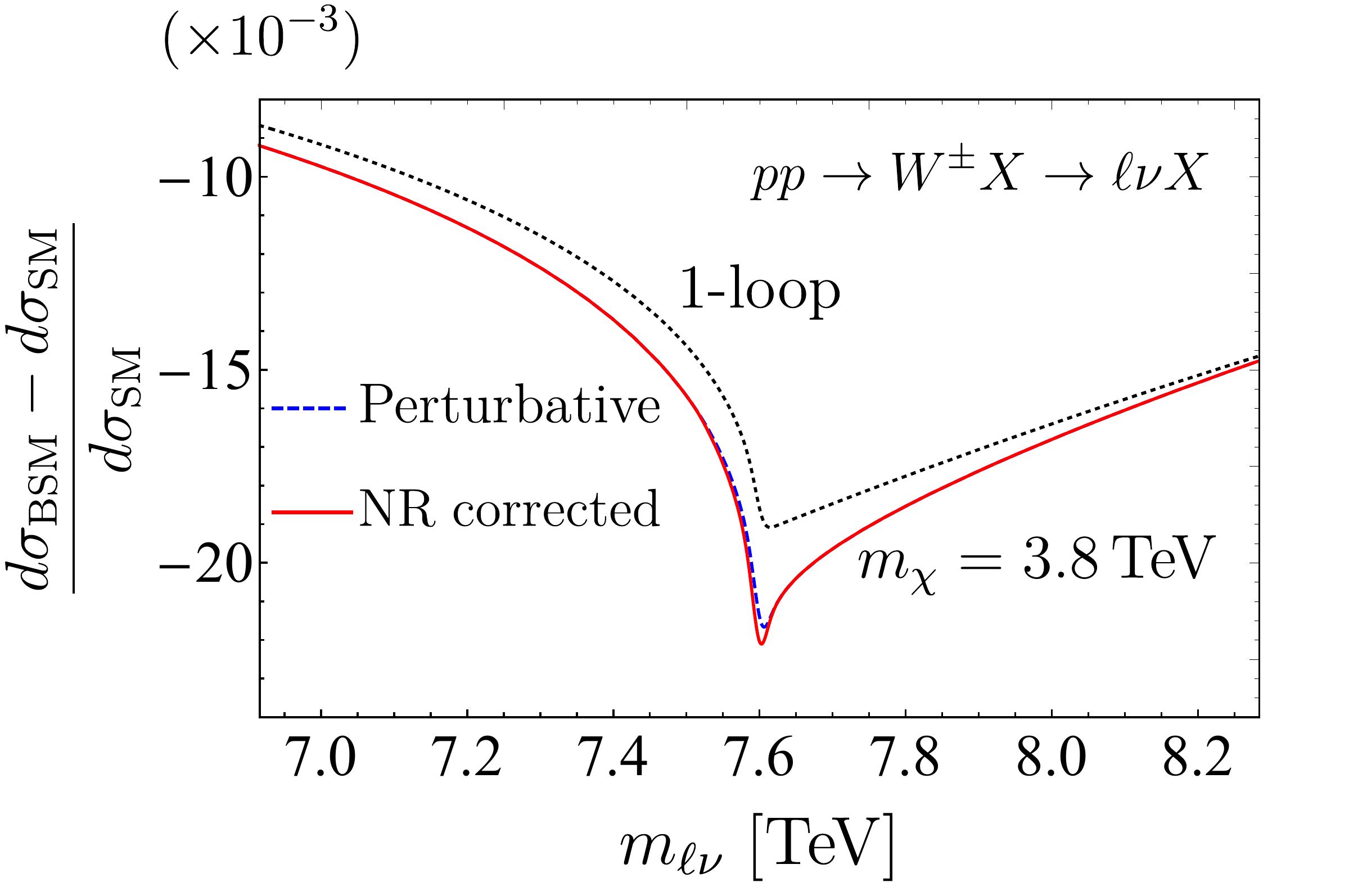}
    \\
    \includegraphics[width=82mm]{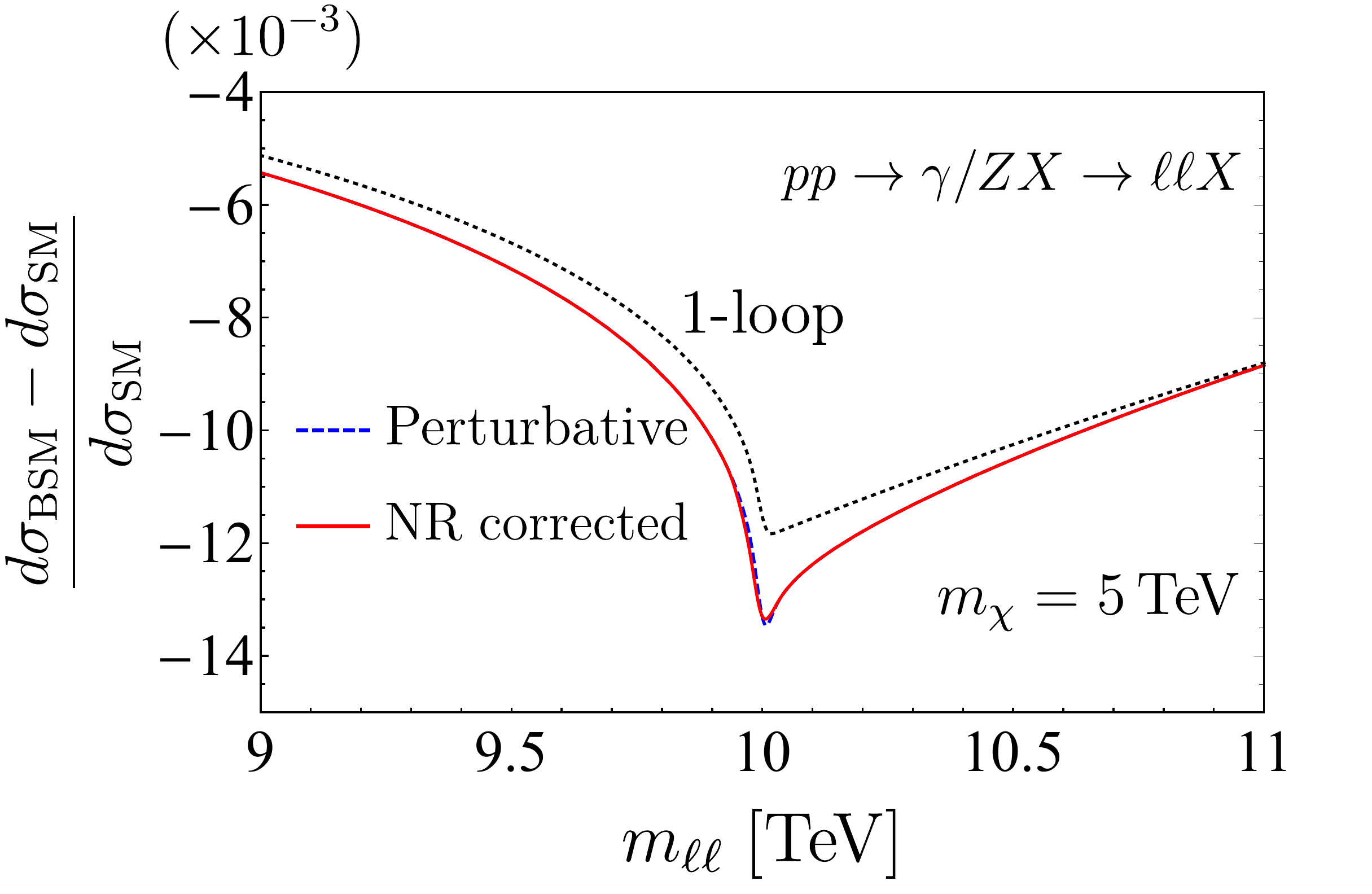}
    \qquad
    \includegraphics[width=82mm]{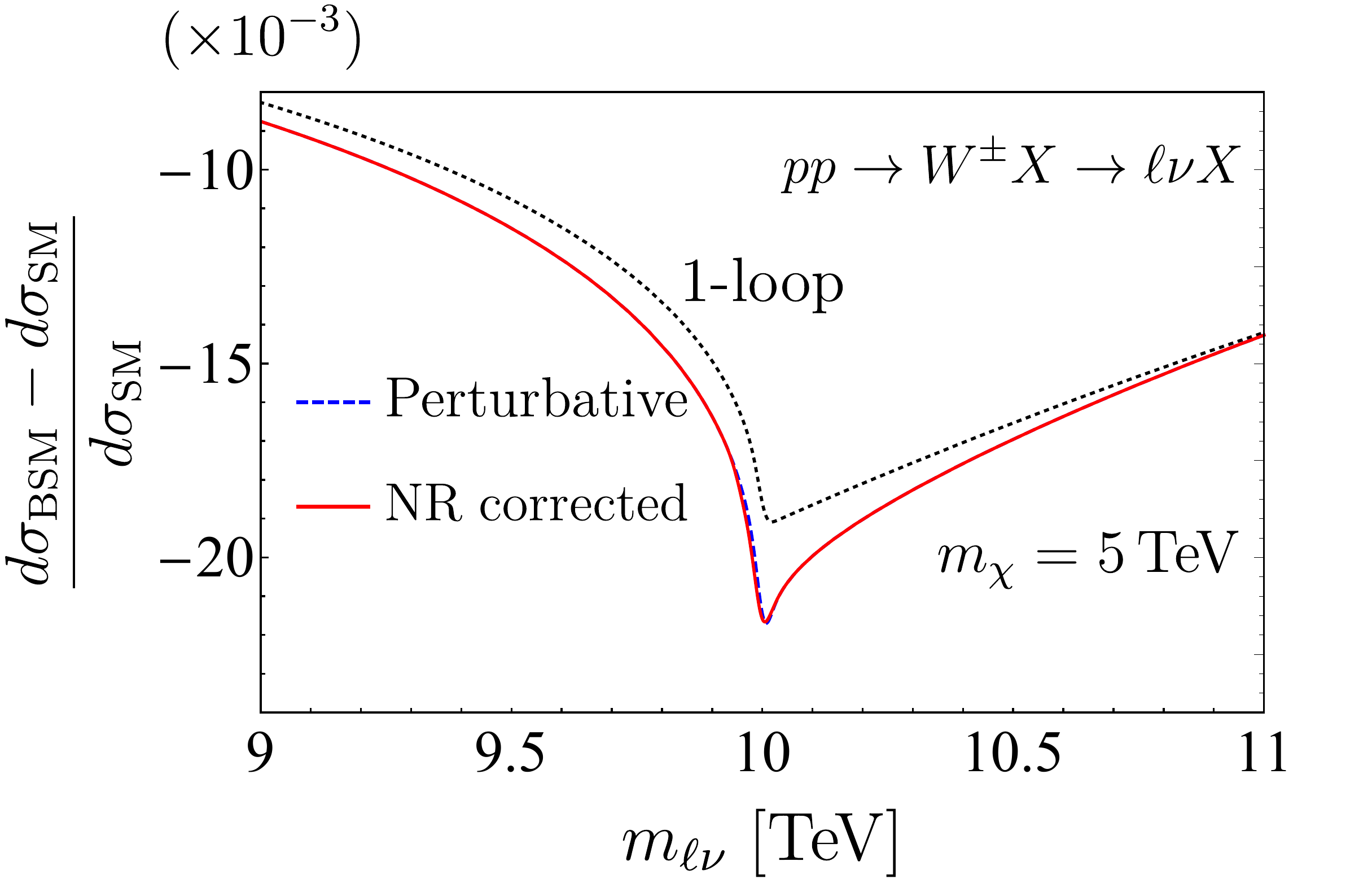}
    \caption{\small The difference between the differential cross sections with and without EWIMP corrections, $(d\sigma_\mathrm{BSM} - d\sigma_\mathrm{SM})/d\sigma_\mathrm{SM}$ for Drell-Yan processes $pp\to \gamma/Z\,X \to \ell \bar{\ell} X$ (left column) and $pp\to W\,X \to \ell \bar{\nu} X$ (right column) are shown as a function of $m_{\ell \ell}$ for  $m_\chi$ is 1 (top row), 3.8 (middle row), and 5\,TeV (bottom low). The energy resolution for the lepton measurement is assumed to be 0.1\,\%. For comparison purposes, the differences obtained by the one-loop calculation (dotted line) and the perturbative calculation, namely one-loop + 2-loop calculation (dashed line) are also shown in each panel.}
    \label{fig: collider signals}
\end{figure}

As we have discussed in the previous section, the matrix element $\mathcal{M}_\mathrm{EWIMP}$ has a divergent property originating in the bound states of the EWIMP. The divergence is regularized by the decay widths of the bound states discussed in the previous section and further smeared by the energy resolution function $f(\delta,\sigma)$ in the calculation of the cross sections. In order to see the effect of the EWIMP quantitatively, we show the difference between the differential cross sections with and without EWIMP contributions in Fig.\,\ref{fig: collider signals} as a function of the lepton invariant mass $m_{\ell \ell}$ for $m_\chi$ = 1, 3.8 and 5\,TeV. As can be seen in all the panels, the two-loop effect gives a certain contribution to the differential cross sections, while the NR effect alters the cross sections slightly in the threshold region. In particular, the effect becomes almost invisible when the EWIMP becomes very heavy, as seen in the bottom panels. When the EWIMP mass is around the one predicting the zero-energy resonance, namely $m_\chi \simeq 4$\,TeV, the NR effect becomes visible as seen in the middle-right panel, though it is still not large enough to detect the effect at the current collider experiment. These results are mainly due to the small electroweak charge of the EWIMP that we are discussing in this article as well as the energy resolution for the lepton measurement. If we think about a EWIMP having a larger electroweak charge and/or future collider experiments having a better energy resolution, the NR effect is expected to be more visible, as we briefly discussed in the next section.

\section{Summary and discussion}

Effects of the EWIMP on Drell-Yan processes for SM lepton productions have been studied in this article. We have estimated one-loop, leading two-loop and non-perturbative contributions to the self-energies of electroweak gauge bosons, and performed the self-consistent matching of those contributions at the threshold region, $q^2 \simeq 4 M_\mathrm{EWIMP}^2$. Then, we have found that the non-perturbative contribution alters the self-energies, thus the differential cross sections of the  Drell-Yan processes significantly, though the contribution from Sommerfeld effects and EWIMP bound states is smeared due to the finite energy resolution for the lepton measurement at the current LHC experiment and does not have a significant impact on the indirect EWIMP detection. On the other hand, we have also found that the higher-order perturbative contribution (the two-loop one) comes with $\alpha^2 \log(M_\mathrm{EWIMP}/m_{Z,W})$ instead of $\alpha^2/\pi^2$ at the threshold region, while the size of the contribution becomes as small as $\alpha^2/\pi^2$ outside the region as expected from the naive loop factor counting. In the triplet EWIMP case, the size of the contribution is $\sim 10$\% of the one-loop contribution in the threshold region, and it gives an important impact for the discovery potential and the measurement of the EWIMP quantum number at current and future hadron collider experiments.

Though we have focused on the case of the triplet Majorana fermion (wino) in this article, our result can be straightforwardly applied to more generic EWIMPs. For the $n$-plet case, the higher-order contribution is expected to be $\sim [(n^5-6n^3+n)/48]\alpha^2  \log(M_\mathrm{EWIMP}/m_{Z,W})$\cite{Machacek:1983tz}, while the one-loop contribution is scaled as $\sim [(n^3-n)/12] \alpha/(4\pi)$. It means that, for $n$ larger than four, the higher-order contribution becomes compatible or larger than the one-loop contribution in the threshold region, and thus the inclusion of the higher-order contribution becomes mandatory for the indirect EWIMP detection at high-energy collider experiments. Moreover, when $n$ is larger, the potential acting between EWIMPs becomes deeper and the EWIMP mass predicting the (first) zero-energy resonance is lighter, and it is then expected to make the non-perturbative contribution such as the Sommerfeld and bound state effects more visible. We remains the study of such higher-order as well as non-perturbative contributions for generic EWIMP cases and of their impacts on collider signals as a future work.

\section*{Acknowledgments}

This work is supported by Grant-in-Aid for Scientific Research from the Ministry of Education, Culture, Sports, Science, and Technology (MEXT), Japan; 17H02878 and 20H01895 (for S.M. \& S.S.), 19H05810 and 20H00153 (for S.M.), 18K13535, 19H04609 and 20H05860 (for S.S.), and by World Premier International Research Center Initiative (WPI), MEXT, Japan.

\bibliographystyle{apsrev4-1}
\bibliography{ref}

\end{document}